\documentclass[longauth]{aa}

\usepackage[varg]{txfonts}

\usepackage{hyperref}
\usepackage{graphicx}
\usepackage{commath}
\usepackage{siunitx}
\usepackage{booktabs}
\usepackage{upgreek}

\usepackage[capitalise]{cleveref} 
\usepackage{csvsimple}
\crefname{figure}{Fig.}{Figs.}
\crefname{section}{Sect.}{Sects.}

\DeclareRobustCommand{\VAN}[3]{#2}
\let\VANthebibliography\thebibliography
\def\thebibliography{\DeclareRobustCommand{\VAN}[3]{##3}\VANthebibliography}

\bibpunct[]{(}{)}{;}{a}{}{,} 

\newcommand{\program}{\textsc}
\newcommand{\ssim}{\sim \!}
\newcommand{\lymana}{{Lyman-\ensuremath{\upalpha}}}
\newcommand{\lymanatext}{\lymana}
\newcommand{\lya}{{Ly\ensuremath{\upalpha}}}
\newcommand{\lyatext}{Lyα}

\newcommand{\HI}{{\ion{H}{I}}}
\newcommand{\HII}{{\ion{H}{II}}}

\newcommand{\OII}{{[\ion{O}{II}]}}
\newcommand{\OIII}{{[\ion{O}{III}]}}
\newcommand{\Halpha}{{\ensuremath{\mathrm{H \upalpha}}}}
\newcommand{\Hbeta}{{\ensuremath{\mathrm{H \upbeta}}}}

\hypersetup{
    unicode=true,
    final=true,
    plainpages=false,
    pdfstartview=FitV,
    pdftoolbar=false,
    pdfmenubar=true,
    bookmarksopen=true,
    bookmarksnumbered=true,
    breaklinks=true,
    linktocpage,
    colorlinks=true,
    linkcolor=blue,
    urlcolor=blue,
    citecolor=blue,
    anchorcolor=green
}
\AtBeginDocument{
  \hypersetup{
    pdftitle={},
    pdfauthor={J. Witstok et al.},
    pdfsubject={},
    pdfkeywords={{Intergalactic medium} -- {}}
  }
}

\begin{document}

\title{Inside the bubble: exploring the environments of reionisation-era \texorpdfstring{\lymana}{\lymanatext} emitting galaxies with JADES\thanks{The main data underlying this study are publicly available on \url{https://archive.stsci.edu/hlsp/jades}.} and FRESCO}
\titlerunning{Exploring the environments of \texorpdfstring{\lymana}{\lymanatext} emitting galaxies}
\subtitle{}

\author{
    Joris Witstok\thanks{E-mail: \href{mailto:jnw30@cam.ac.uk}{jnw30@cam.ac.uk}}\inst{\ref{inst:Kavli}}\fnmsep\inst{\ref{inst:Cav}}
    \and Renske Smit\inst{\ref{inst:LJMU}}
    \and Aayush Saxena\inst{\ref{inst:Oxford}}\fnmsep\inst{\ref{inst:UCL}}
    \and Gareth C. Jones\inst{\ref{inst:Oxford}}
    \and Jakob M. Helton\inst{\ref{inst:Steward}}
    \and Fengwu Sun\inst{\ref{inst:Steward}}
    \and Roberto Maiolino\inst{\ref{inst:Kavli}}\fnmsep\inst{\ref{inst:Cav}}\fnmsep\inst{\ref{inst:UCL}}
    \and Nimisha Kumari\inst{\ref{inst:AURA}}
    \and Daniel P. Stark\inst{\ref{inst:Steward}}
    \and Andrew J.\ Bunker\inst{\ref{inst:Oxford}}
    \and Santiago Arribas\inst{\ref{inst:CAB}}
    \and William M. Baker\inst{\ref{inst:Kavli}}\fnmsep\inst{\ref{inst:Cav}}
    \and Rachana Bhatawdekar\inst{\ref{inst:ESAC}}\fnmsep\inst{\ref{inst:ESTEC}}
    \and Kristan Boyett\inst{\ref{inst:Melbourne}}\fnmsep\inst{\ref{inst:ASTRO3D}}
    \and Alex J. Cameron\inst{\ref{inst:Oxford}}
    \and Stefano Carniani\inst{\ref{inst:SNS}}
    \and Stephane Charlot\inst{\ref{inst:IAP}}
    \and Jacopo Chevallard\inst{\ref{inst:Oxford}}
    \and Mirko Curti\inst{\ref{inst:ESO}}\fnmsep\inst{\ref{inst:Kavli}}\fnmsep\inst{\ref{inst:Cav}}
    \and Emma Curtis-Lake\inst{\ref{inst:Hertfordshire}}
    \and Daniel J.\ Eisenstein\inst{\ref{inst:CfA}}
    \and Ryan Endsley\inst{\ref{inst:UoT}}
    \and Kevin Hainline\inst{\ref{inst:Steward}}
    \and Zhiyuan Ji\inst{\ref{inst:Steward}}
    \and Benjamin D.\ Johnson\inst{\ref{inst:CfA}}
    \and Tobias J. Looser\inst{\ref{inst:Kavli}}\fnmsep\inst{\ref{inst:Cav}}
    \and Erica Nelson\inst{\ref{inst:UoC}}
    \and Michele Perna\inst{\ref{inst:CAB}}
    \and Hans-Walter Rix\inst{\ref{inst:MPIA}}
    \and Brant E. Robertson\inst{\ref{inst:UCSC}}
    \and Lester Sandles\inst{\ref{inst:Kavli}}\fnmsep\inst{\ref{inst:Cav}}
    \and Jan Scholtz\inst{\ref{inst:Kavli}}\fnmsep\inst{\ref{inst:Cav}}
    \and Charlotte Simmonds\inst{\ref{inst:Kavli}}\fnmsep\inst{\ref{inst:Cav}}
    \and Sandro Tacchella\inst{\ref{inst:Kavli}}\fnmsep\inst{\ref{inst:Cav}}
    \and Hannah \"Ubler\inst{\ref{inst:Kavli}}\fnmsep\inst{\ref{inst:Cav}}
    \and Christina C. Williams\inst{\ref{inst:NOIRLab}}
    \and Christopher N. A. Willmer\inst{\ref{inst:Steward}}
    \and Chris Willott\inst{\ref{inst:NRC}}
}
\authorrunning{J. Witstok et al.}

\institute{
    Kavli Institute for Cosmology, University of Cambridge, Madingley Road, Cambridge CB3 0HA, UK\label{inst:Kavli}
    \and Cavendish Laboratory, University of Cambridge, 19 JJ Thomson Avenue, Cambridge CB3 0HE, UK\label{inst:Cav}
    \and Astrophysics Research Institute, Liverpool John Moores University, 146 Brownlow Hill, Liverpool L3 5RF, UK\label{inst:LJMU}
    \and Department of Physics, University of Oxford, Denys Wilkinson Building, Keble Road, Oxford OX1 3RH, UK\label{inst:Oxford}
    \and Department of Physics and Astronomy, University College London, Gower Street, London WC1E 6BT, UK\label{inst:UCL}
    \and Steward Observatory, University of Arizona, 933 N. Cherry Ave., Tucson, AZ 85721, USA\label{inst:Steward}
    \and AURA for European Space Agency, Space Telescope Science Institute, 3700 San Martin Drive, Baltimore, MD 21210, USA\label{inst:AURA}
    \and Centro de Astrobiolog\'ia (CAB), CSIC–INTA, Cra. de Ajalvir Km.~4, 28850- Torrej\'on de Ardoz, Madrid, Spain\label{inst:CAB}
    \and European Space Agency (ESA), European Space Astronomy Centre (ESAC), Camino Bajo del Castillo s/n, 28692 Villanueva de la Cañada, Madrid, Spain\label{inst:ESAC}
    \and European Space Agency, ESA/ESTEC, Keplerlaan 1, 2201 AZ Noordwijk, NL\label{inst:ESTEC}
    \and School of Physics, University of Melbourne, Parkville 3010, VIC, Australia\label{inst:Melbourne}
    \and ARC Centre of Excellence for All Sky Astrophysics in 3 Dimensions (ASTRO 3D), Australia\label{inst:ASTRO3D}
    \and Scuola Normale Superiore, Piazza dei Cavalieri 7, I-56126 Pisa, Italy\label{inst:SNS}
    \and Sorbonne Universit\'e, CNRS, UMR 7095, Institut d'Astrophysique de Paris, 98 bis bd Arago, 75014 Paris, France\label{inst:IAP}
    \and European Southern Observatory, Karl-Schwarzschild-Strasse 2, 85748 Garching, Germany\label{inst:ESO}
    \and Centre for Astrophysics Research, Department of Physics, Astronomy and Mathematics, University of Hertfordshire, Hatfield AL10 9AB, UK\label{inst:Hertfordshire}
    \and Center for Astrophysics $|$ Harvard \& Smithsonian, 60 Garden St., Cambridge MA 02138, USA\label{inst:CfA}
    \and Department of Astronomy, University of Texas, Austin TX 78712, USA\label{inst:UoT}
    \and Department for Astrophysical and Planetary Science, University of Colorado, Boulder, CO 80309, USA\label{inst:UoC}
    \and Max-Planck-Institut f\"ur Astronomie, K\"onigstuhl 17, D-69117, Heidelberg, Germany\label{inst:MPIA}
    \and Department of Astronomy and Astrophysics University of California, Santa Cruz, 1156 High Street, Santa Cruz, CA 96054, USA\label{inst:UCSC}
    \and NSF's National Optical-Infrared Astronomy Research Laboratory, 950 North Cherry Avenue, Tucson, AZ 85719, USA\label{inst:NOIRLab}
    \and NRC Herzberg, 5071 West Saanich Rd, Victoria, BC V9E 2E7, Canada\label{inst:NRC}
}


\abstract{
    We present a study of the environments of $17$ \lymana\ emitting galaxies (LAEs) in the reionisation era ($5.8 < z < 8$) identified by \textit{JWST}/NIRSpec as part of the \textit{JWST} Advanced Deep Extragalactic Survey (JADES). Unless situated in sufficiently (re)ionised regions, \lymana\ emission from these galaxies would be strongly absorbed by neutral gas in the intergalactic medium (IGM). We conservatively estimate sizes of the ionised regions required to reconcile the relatively low \lymana\ velocity offsets ($\Delta v_\text{\lya} < 300 \, \mathrm{km \, s^{-1}}$) with moderately high \lymana\ escape fractions ($f_\text{esc, \lya} > 5\%$) observed in our sample of LAEs, suggesting the presence of ionised hydrogen along the line of sight towards at least eight out of $17$ LAEs. We find minimum physical `bubble' sizes of the order of $R_\text{ion} \sim 0.1$-$1 \, \mathrm{pMpc}$ are required in a patchy reionisation scenario where ionised bubbles containing the LAEs are embedded in a fully neutral IGM. Around half of the LAEs in our sample are found to coincide with large-scale galaxy overdensities seen in FRESCO at $z \sim 5.8$-$5.9$ and $z \sim 7.3$, suggesting \lymana\ transmission is strongly enhanced in such overdense regions, and underlining the importance of LAEs as tracers of the first large-scale ionised bubbles. Considering only spectroscopically confirmed galaxies, we find our sample of UV-faint LAEs ($M_\text{UV} \gtrsim -20 \, \mathrm{mag}$) and their direct neighbours are generally not able to produce the required ionised regions based on the \lymana\ transmission properties, suggesting lower-luminosity sources likely play an important role in carving out these bubbles. These observations demonstrate the combined power of \textit{JWST} multi-object and slitless spectroscopy in acquiring a unique view of the early Universe during cosmic reionisation via the most distant LAEs.
}


\maketitle

\section{Introduction}
\label{sec:Introduction}

Cosmic Dawn marked the formation of the first astrophysical objects a few hundred million years after the Big Bang ($z \gtrsim 15$), bringing the Dark Ages to an end and, importantly, setting in motion the process of cosmic reionisation \citep{2018PhR...780....1D}. Optical-depth measurements of the cosmic microwave background indicate reionisation reached a midpoint at approximately $z \sim 8$, suggesting that this momentous phase transition transpired broadly in the first billion years of cosmic time \citep[e.g.][]{2020A&A...641A...6P}. In the past decade, various independent methods tracing the neutrality of the intergalactic medium (IGM) have started to further elucidate the timeline and origin of reionisation, indicating the bulk of neutral hydrogen in the Universe was reionised at $6 \lesssim z \lesssim 10$, likely driven by the ionising radiation of star-forming galaxies \citep{2015ApJ...802L..19R, 2019ApJ...879...36F, 2020ApJ...892..109N}. One powerful probe of the ionising capabilities of reionisation-era galaxies is \lymana\ (\lya), one of the most prominent electronic transitions of hydrogen \citep{1967ApJ...147..868P}.

Unless the IGM is highly ionised, photons escaping a galaxy at a rest-frame wavelength of $\lambda_\text{emit} \lesssim \lambda_\text{\lya} = 1215.67 \, \mathrm{\AA}$ are nearly entirely scattered out of the line of sight before they redshift away from resonance \citep{1965ApJ...142.1633G}. If there is little to no significant evolution in the intrinsic \lya\ properties of galaxies, reionisation should therefore be characterised by a rapid decline towards higher redshift of the fraction of galaxies observed to have \lymana\ emission \citep{2004MNRAS.354..695F, 2014PASA...31...40D, 2020ARA&A..58..617O}. Indeed, a plummeting fraction of \lymana\ emitting galaxies (LAEs) has been interpreted as evidence of a rapidly evolving neutral hydrogen fraction between $z \sim 6$ and $z \sim 8$ \citep{2010MNRAS.408.1628S, 2017MNRAS.464..469S, 2011ApJ...743..132P, 2012ApJ...744...83O, 2012MNRAS.427.3055C, 2014MNRAS.443.2831C, 2013ApJ...775L..29T, 2014ApJ...795...20S, 2014ApJ...794....5T, 2018ApJ...856....2M, 2019MNRAS.485.3947M, 2020ApJ...891L..10T, 2020MNRAS.495.3602W}.

Beyond providing global IGM constraints, however, LAEs can be exploited to trace the detailed progression of reionisation. Reionisation has been shown to be an inhomogeneous process \citep[e.g.][]{2014ApJ...793..113P, 2018ApJ...863...92B, 2019MNRAS.485L..24K, 2020MNRAS.491.1736K, 2022MNRAS.514...55B}, suggesting that galaxies in overdense regions are expected to carve out the first ionised `bubbles' that preferentially allow \lya\ emission to be observed when they have grown sufficiently large \citep[e.g.][]{2018MNRAS.479.2564W, 2018ApJ...857L..11M, 2021MNRAS.502.6044E, 2022MNRAS.517.5642E, 2022MNRAS.510.3858Q, 2022arXiv221209850J, 2022ApJ...933...87J, 2022MNRAS.515.5790L}. The most distant LAEs therefore form a crucial observational frontier in mapping the reionisation process, providing unique signposts of the first large-scale, ionised patches of the Universe \citep{2021NatAs...5..485H, 2022MNRAS.511.6042E, 2022ApJ...930..104L, 2023arXiv230411192L, 2023MNRAS.524.5891T, 2023arXiv230516670W}.

Building on the legacy of the \textit{Hubble Space Telescope} (\textit{HST}), the successful commissioning of \textit{JWST} \citep{2023PASP..135e8001M, 2023PASP..135d8001R}, specifically designed for finding and characterising the first generation of galaxies \citep{2022ARA&A..60..121R, 2023PASP..135f8001G}, now places us in prime position to address several fundamental questions surrounding reionisation. In particular, with spectroscopic coverage of the rest-frame UV and optical, \textit{JWST} allows for the unprecedented characterisation of LAEs throughout the epoch of reionisation (EoR) for the first time \citep{2023MNRAS.526.1657T, 2023arXiv230405385J, 2023arXiv230316225W, 2023arXiv230516670W}. Crucially, the \textit{JWST}/Near-Infrared Spectrograph \citep[NIRSpec;][]{2022A&A...661A..80J, 2023PASP..135c8001B} possesses the required sensitivity, and particularly the multiplexing capabilities \citep[via its micro-shutter assembly or MSA;][]{2022A&A...661A..81F}, to extend pioneering ground-based efforts that were restricted to the few brightest sources \citep[e.g.][]{2012ApJ...744...83O, 2013Natur.502..524F, 2015ApJ...810L..12Z, 2015ApJ...804L..30O, 2016ApJ...823..143R}. Reaching absolute AB magnitudes in the UV of $M_\text{UV} \sim -17 \, \mathrm{mag}$ even without gravitational lensing, \textit{JWST} finally enables an extensive investigation of the faint-end of the luminosity function to reveal which sources are the dominant drivers of cosmic reionisation \citep[e.g.][]{2023A&A...678A..68S}.

\begin{table*}[h]
    \centering
    \footnotesize
    \caption{Observed LAE properties.}
    \begin{tabular}{llllllll}
        ID & JADES source name & $z_\text{spec}$ & $M_\text{UV} \, (\mathrm{mag})$ & $\beta_\text{UV}$ & $\text{EW}_\text{\lya} \, (\mathrm{\AA})$ & $\Delta v_\text{\lya} \, (\mathrm{km \, s^{-1}})$ & $f_\text{esc, \lya}$ \\
        \midrule
        \begin{filecontents*}{Bubble_sizes.csv}
NID,ID,IDfoot,MID,RA,Dec,zspec,MUV,betaUV,EW,dv,fesc,Rion,RiongxHI,RionLAE,RionLAEgxHI,Riontot,RiontotgxHI
10056849,JADES-GS+53.11351-27.77284,JADES-GS+53.11351-27.77284,,$+53.11351$,$-27.77284$,$5.814$,$-18.14 \pm 0.04$,$-2.49 \pm 0.04$,$97 \pm 15$,$215$,$0.29 \pm 0.04$,$0.52$,\dots,$0.18$,\dots,$0.18$,\dots
19606,JADES-GS+53.17655-27.77111,JADES-GS+53.17655-27.77111\tablefootmark{$\ast$},3203,$+53.17655$,$-27.77111$,$5.889$,$-18.61 \pm 0.03$,$-2.70 \pm 0.06$,$89 \pm 11$,$53$\tablefootmark{$\ast\ast$},$0.36 \pm 0.03$,$0.93$,\dots,$0.18$,\dots,$0.34$,\dots
9365,JADES-GS+53.16280-27.76084,JADES-GS+53.16280-27.76084\tablefootmark{$\ast$},3089,$+53.16280$,$-27.76084$,$5.917$,$-19.37 \pm 0.01$,$-2.52 \pm 0.09$,$118 \pm 28$,$180$,$0.28 \pm 0.04$,$0.55$,\dots,$0.19$,\dots,$0.26$,\dots
9422,JADES-GS+53.12175-27.79763,JADES-GS+53.12175-27.79763,,$+53.12175$,$-27.79763$,$5.937$,$-19.80 \pm 0.01$,$-2.33 \pm 0.04$,$124 \pm 15$,$175$,$0.26 \pm 0.01$,$0.52$,\dots,$0.26$,\dots,$0.47$,\dots
6002,JADES-GS+53.11041-27.80892,JADES-GS+53.11041-27.80892,,$+53.11041$,$-27.80892$,$5.937$,$-18.72 \pm 0.04$,$-2.59 \pm 0.01$,$50.5 \pm 5.8$,$170$,$0.35 \pm 0.04$,$0.74$,\dots,$0.14$,\dots,$0.48$,\dots
19342,JADES-GS+53.16062-27.77161,JADES-GS+53.16062-27.77161\tablefootmark{$\ast$},547,$+53.16062$,$-27.77161$,$5.974$,$-18.55 \pm 0.05$,$-2.75 \pm 0.04$,$49.9 \pm 9.6$,$279$,$0.24 \pm 0.04$,$0.33$,\dots,$0.15$,\dots,$0.20$,\dots
17138,JADES-GS+53.08604-27.74760,JADES-GS+53.08604-27.74760,,$+53.08604$,$-27.74760$,$6.204$,$-19.34 \pm 0.04$,$-2.26 \pm 0.54$,$94 \pm 41$,$0$,$0.40 \pm 0.10$,$1.13$,$0.16$,$0.15$,$0.15$,$0.15$,$0.15$
58850,JADES-GS+53.09517-27.76061,JADES-GS+53.09517-27.76061,,$+53.09517$,$-27.76061$,$6.263$,$-19.82 \pm 0.03$,$-1.93 \pm 0.06$,$16.3 \pm 3.9$,$230$,$0.07 \pm 0.02$,$0.08$,\dots,$0.25$,\dots,$0.25$,\dots
14123,JADES-GS+53.17836-27.80098,JADES-GS+53.17836-27.80098\tablefootmark{$\ast$},6231,$+53.17836$,$-27.80098$,$6.327$,$-19.20 \pm 0.03$,$-2.26 \pm 0.21$,$150 \pm 100$,$106$,$0.35 \pm 0.07$,$0.85$,$0.08$,$0.19$,$0.19$,$0.31$,$0.19$
18846,JADES-GS+53.13492-27.77271,JADES-GS+53.13492-27.77271,,$+53.13492$,$-27.77271$,$6.336$,$-19.90 \pm 0.01$,$-2.43 \pm 0.01$,$44.2 \pm 1.7$,$114$,$0.26 \pm 0.01$,$0.62$,$0.02$,$0.20$,$0.20$,$0.22$,$0.20$
13607,JADES-GS+53.13743-27.76519,JADES-GS+53.13743-27.76519,,$+53.13743$,$-27.76519$,$6.622$,$-19.37 \pm 0.02$,$-1.79 \pm 0.29$,$33 \pm 11$,$128$,$0.26 \pm 0.08$,$0.61$,$0.15$,$0.18$,$0.18$,$0.20$,$0.18$
16625,JADES-GS+53.16904-27.77884,JADES-GS+53.16904-27.77884\tablefootmark{$\ast$},852,$+53.16904$,$-27.77884$,$6.631$,$-18.60 \pm 0.04$,$-2.59 \pm 0.02$,$51.0 \pm 7.4$,$242$,$0.14 \pm 0.02$,$0.23$,\dots,$0.16$,\dots,$0.20$,\dots
4297,JADES-GS+53.15579-27.81520,JADES-GS+53.15579-27.81520,,$+53.15579$,$-27.81520$,$6.712$,$-18.48 \pm 0.04$,$-2.39 \pm 0.09$,$106 \pm 23$,$153$,$0.55 \pm 0.04$,$1.53$,$0.61$,$0.17$,$0.17$,$0.19$,$0.19$
15362,JADES-GS+53.11634-27.76194,JADES-GS+53.11634-27.76194,,$+53.11634$,$-27.76194$,$6.794$,$-18.86 \pm 0.26$,$-2.14 \pm 0.15$,$50 \pm 28$,$27$,$0.20 \pm 0.07$,$0.62$,$0.30$,$0.14$,$0.14$,$0.14$,$0.14$
10013682,JADES-GS+53.16746-27.77201,JADES-GS+53.16746-27.77201\tablefootmark{$\dagger$},,$+53.16746$,$-27.77201$,$7.276$,$-16.86 \pm 0.28$,$-2.17 \pm 0.60$,$337 \pm 175$,$178$,$0.67 \pm 0.18$,$2.35$,$1.62$,$0.099$,$0.099$,$0.41$,$0.37$
12637,JADES-GS+53.13347-27.76037,JADES-GS+53.13347-27.76037\tablefootmark{$\ddagger$},,$+53.13347$,$-27.76037$,$7.66$,$-20.59 \pm 0.07$,$-2.20 \pm 0.02$,$24.0 \pm 1.9$,$131$,$0.15 \pm 0.01$,$0.42$,$0.31$,$0.23$,$0.23$,$0.23$,$0.23$
21842,JADES-GS+53.15682-27.76716,JADES-GS+53.15682-27.76716,,$+53.15682$,$-27.76716$,$7.98$,$-18.80 \pm 0.06$,$-2.52 \pm 0.03$,$29.2 \pm 3.3$,$84$,$0.09 \pm 0.01$,$0.35$,$0.30$,$0.14$,$0.14$,$0.14$,$0.14$
\end{filecontents*}
\csvreader[late after line=\\, head to column names]{Bubble_sizes.csv}{}{\NID & \IDfoot & \zspec & \MUV & \betaUV & \EW & \dv & \fesc}
        \bottomrule
    \end{tabular}
    \tablefoot{
        Listed properties of each LAE are its NIRSpec ID \citetext{as presented in Bunker et al. 2023b}, full JADES identifier (including their J2000 Right Ascension and Declination in $\mathrm{deg}$), spectroscopic redshift ($z_\text{spec}$), absolute magnitude in the UV ($M_\text{UV}$), UV spectral slope ($\beta_\text{UV}$), and the \lya\ properties measured in the R1000 spectra \citep{2023arXiv230604536S}: rest-frame EW ($\text{EW}_\text{\lya}$), velocity offset ($\Delta v_\text{\lya}$; carrying an uncertainty of $\ssim 100 \, \mathrm{km \, s^{-1}}$ as discussed in \cref{sssec:Data_analysis}), and escape fraction ($f_\text{esc, \lya}$). \\
        \tablefoottext{$\ast$}{Contained within the MUSE HUDF surveys \citep[; see \cref{ssec:MUSE_observations}]{2023A&A...670A...4B}.} \\
        \tablefoottext{$\ast\ast$}{\lya\ velocity offset adopted from MUSE HUDF surveys \citep{2023A&A...670A...4B}.} \\
        \tablefoottext{$\dagger$}{JADES-GS-z7-LA \citep{2023A&A...678A..68S}.} \\
        \tablefoottext{$\ddagger$}{z7-GSD-3811 \citep{2016ApJ...826..113S}.}
    }
    \label{tab:LAE_properties}
\end{table*}
\begin{figure}
	\centering
	\includegraphics[width=\linewidth]{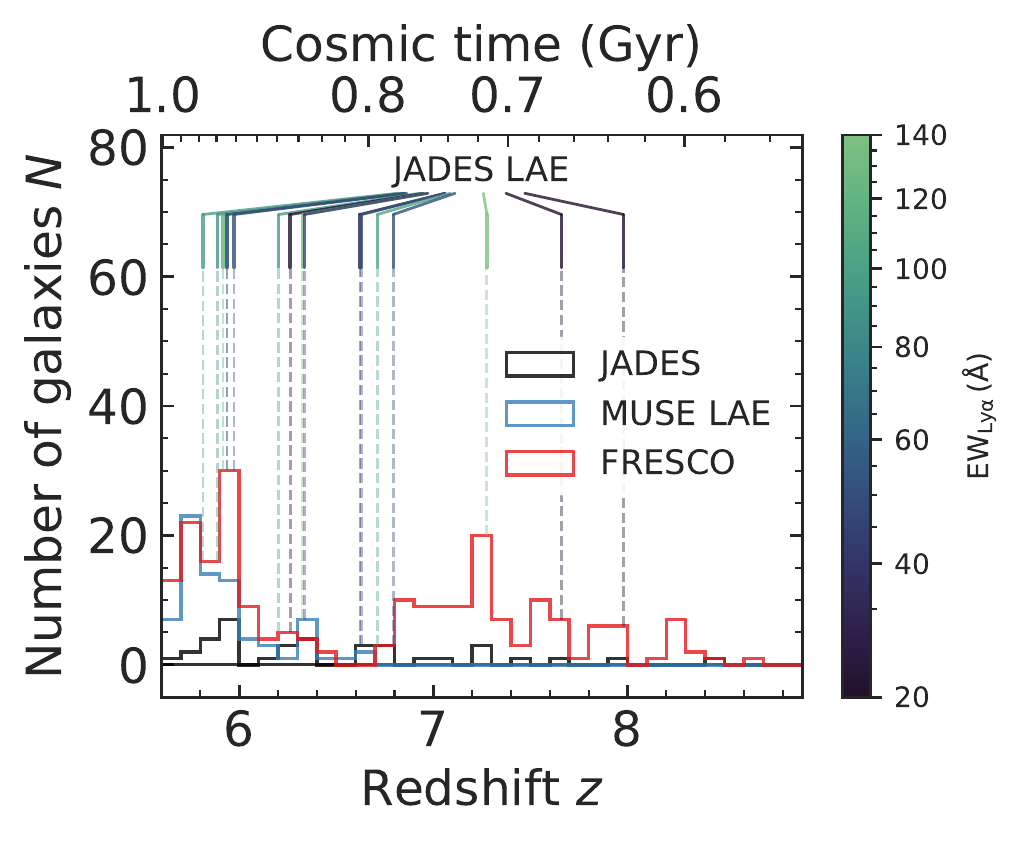}
	\caption{Redshift histogram of spectroscopically confirmed galaxies over GOODS-S. Sources confirmed in JADES NIRSpec observations (\cref{ssec:JADES_observations}) are shown by the black histogram \citetext{Bunker et al. 2023c}, while the red distribution is of those identified in FRESCO data \citep{2023MNRAS.525.2864O}, discussed in \cref{ssec:FRESCO_observations}. EoR ($z \gtrsim 5.8$) LAEs identified in JADES \citep{2023arXiv230602471J, 2023arXiv230604536S} are marked by vertical solid-dashed lines, coloured by \lya\ rest-frame EW according to the scale on the right. Additional LAEs found at $z > 5.5$ in the MUSE HUDF surveys (\citealt{2023A&A...670A...4B}), discussed in \cref{ssec:MUSE_observations}, are shown by the light blue distribution.
	}
	\label{fig:LAE_redshift_distribution}
\end{figure}
\begin{figure*}
	\centering
	\includegraphics[width=0.9\linewidth]{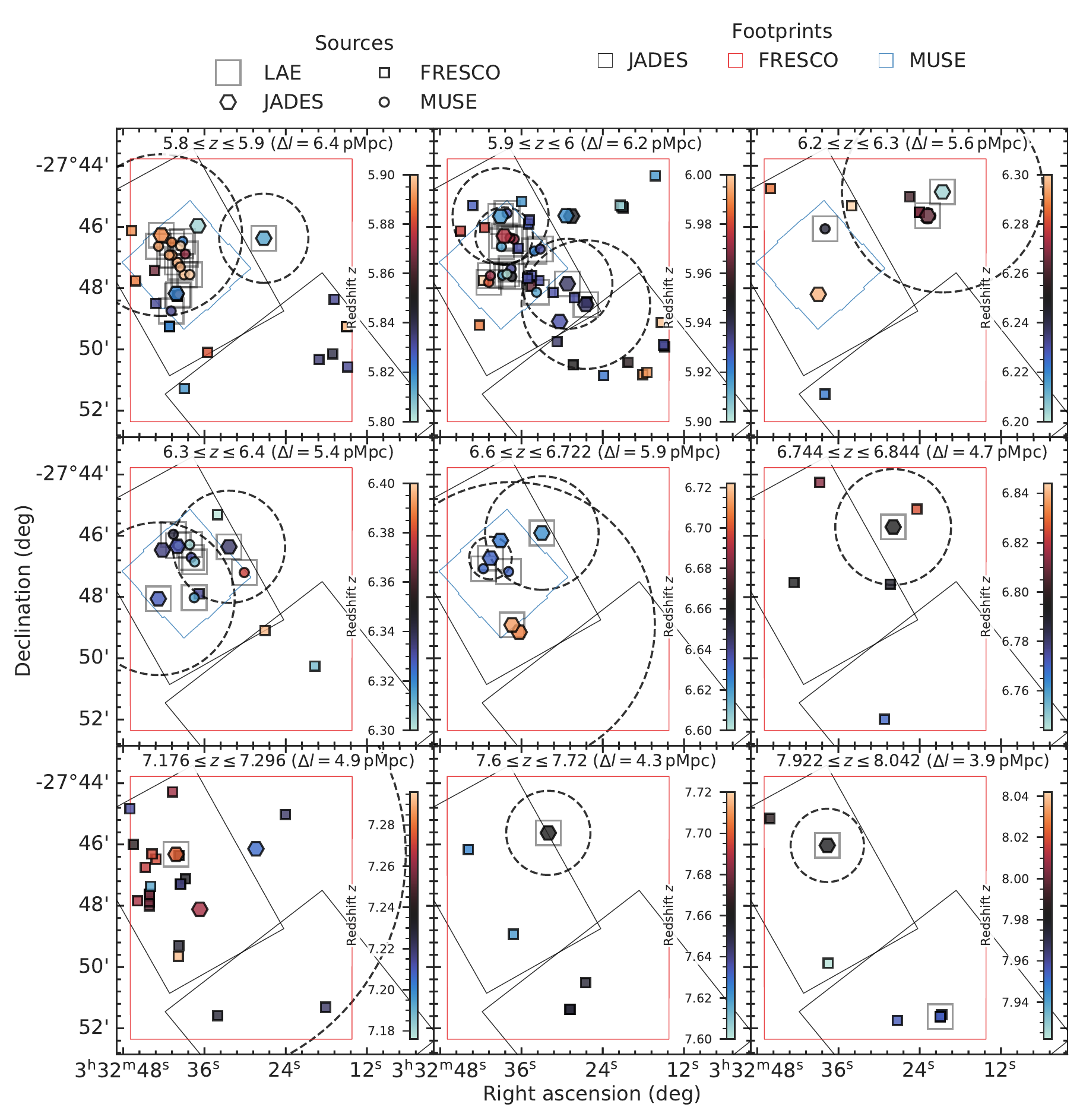}
	\caption{On-sky distribution of spectroscopically confirmed galaxies in the vicinity of $z \gtrsim 5.8$ LAEs identified in JADES NIRSpec observations over GOODS-S. Sources found in the JADES, FRESCO, and MUSE HUDF surveys are respectively shown by hexagons, squares, and circles; larger grey squares indicate LAEs. Each panel shows a different redshift slice, approximately corresponding to a distance along the line of sight between $2 \, \mathrm{pMpc} \lesssim \Delta l \lesssim 6 \, \mathrm{pMpc}$. The JADES footprint is indicated by the black rectangles, FRESCO is highlighted in red, and the total (MOSAIC) extent of the MUSE HUDF surveys is shown by the light blue rectangle. Projected sizes of the ionised bubbles centred on each LAE, inferred based on the observed \lya\ properties (with $\bar{x}_\text{\HI} = 1$; see \cref{ssec:Ionised_bubble_sizes}), are shown as dashed circles.
	}
	\label{fig:LAE_on-sky_distribution}
\end{figure*}

Here, we aim to investigate the environments, including potential ionised bubbles, of such faint ($M_\text{UV} \gtrsim -20 \, \mathrm{mag}$) reionisation-era galaxies which have been observed to exhibit \lya\ emission in recent \textit{JWST} measurements \citep{2023arXiv230602471J}. We restrict this current study to redshifts of $z \gtrsim 5.8$, considering the global neutral hydrogen fraction approaches $\bar{x}_\text{\HI} \sim 0$ at later times even in `late' reionisation scenarios \citep[e.g.][]{2018MNRAS.479.2564W, 2019MNRAS.485.1350W, 2022ApJ...932...76Z}. The outline of this work is as follows: in \cref{sec:Observations}, we discuss the data set considered in this work. \Cref{sec:Results_and_discussion} presents our findings on the potential ionised bubbles surrounding these galaxies and discusses these findings in context of previous works, while \cref{sec:Summary_and_conclusions} summarises our results. We consistently use a flat $\Lambda$CDM cosmology based on the latest results of the Planck collaboration (i.e. $H_0 = 67.4 \, \mathrm{km \, s^{-1} \, Mpc^{-1}}$, $\Omega_\text{m} = 0.315$, $\Omega_\text{b} = 0.0492$; \citealt{2020A&A...641A...6P}) and a cosmic hydrogen fraction of $f_\text{H} = 0.76$ throughout. On-sky separations of $1\arcsec$ and $1\arcmin$ at $z = 7$ hence correspond to $5.34 \, \text{physical kpc}$ (pkpc) and $0.32 \, \text{physical Mpc}$ (pMpc), respectively. Quoted magnitudes are in the AB system \citep{1983ApJ...266..713O}.

\section{Observations}
\label{sec:Observations}

In the following sections, we discuss several spectroscopic data sets containing both the main LAEs considered in this work as well as the reference sample of galaxies used to characterise their environments. We show their redshift histograms in \cref{fig:LAE_redshift_distribution}, while \cref{fig:LAE_on-sky_distribution} shows their distribution on the sky.

\subsection{JADES}
\label{ssec:JADES_observations}

\subsubsection{Data and sample selection}
\label{sssec:Data_and_sample_selection}

The data presented in this work were obtained as part of the DEEP and MEDIUM tiers of the \textit{JWST} Advanced Deep Extragalactic Survey \citep[JADES;][]{2023arXiv230602465E}, a joint survey conducted by the guaranteed time observations (GTO) instrument science teams of the \textit{JWST}/Near-Infrared Camera \citep[NIRCam;][]{2023PASP..135b8001R} and \textit{JWST}/NIRSpec. Specifically, deep NIRCam imaging was taken over the Great Observatories Origins Deep Survey – South \citep[GOODS-S;][]{2004ApJ...600L..93G}, containing the Hubble Ultra Deep Field \citep[HUDF; e.g.][]{2006AJ....132.1729B} previous JADES works \citep[e.g.][]{2023NatAs...7..611R}. The imaging (\textit{JWST} programme 1180; PI: Eisenstein) spans $65 \, \mathrm{arcmin^2}$ over a wavelength range of $\lambda_\text{obs} \simeq 0.8 \, \mathrm{\upmu m}$ to $5 \, \mathrm{\upmu m}$ and reaches $m_\text{F200W} \sim 30 \, \mathrm{mag}$ \citep{2023ApJS..269...16R}. This is complemented by public medium-band imaging from the \textit{JWST} Extragalactic Medium-band Survey (JEMS) in the HUDF \citep[\textit{JWST} programme 1963, PI: Williams;][]{2023ApJS..268...64W} and from the First Reionization Epoch Spectroscopic COmplete Survey \citep[FRESCO; \textit{JWST} programme 1895, PI: Oesch;][]{2023MNRAS.525.2864O}.

Multi-object spectroscopy in GOODS-S was subsequently obtained in the NIRSpec MSA mode \citep{2022A&A...661A..81F}, primarily targeting high-redshift galaxy candidates selected in a combination of NIRCam and publicly available \textit{HST} imaging \citep{2023arXiv230602467B}. Under \textit{JWST} programme 1210 (PI: L\"{u}tzgendorf), targets were observed both in the PRISM/CLEAR spectral configuration (simply PRISM hereafter; spectral range $0.6 \, \mathrm{\upmu m}$ to $5.3 \, \mathrm{\upmu m}$, resolving power $R \sim 100$) and `R1000' medium-resolution gratings (in grating-filter combinations G140M/F070LP, G235M/F170LP, and G395M/F290LP; resolving power $R \sim 1000$; e.g. \citealt{2023A&A...677A.115C}; \citealt{2023arXiv230408516C}). Total exposure times in the DEEP tier ranged from 9.3 to 28 hours \citep[depending on slit placement across three visits; e.g.][]{2023NatAs...7..622C} in the PRISM configuration, with the R1000 gratings each receiving a third of the PRISM integration time, whereas the MEDIUM tier targeted a larger number of sources at shallower depth \citep[between $1$ and $2$ hours for both PRISM and R1000 configurations;][]{2023arXiv230602471J}.

\subsubsection{Data reduction}
\label{sssec:Data_reduction}

The data reduction procedure of the NIRSpec observations have been outlined in preceding JADES works \citep[e.g.][]{2023NatAs...7..622C, 2023arXiv230602465E} and will be described in detail in a forthcoming paper \citetext{Carniani \& NIRSpec GTO collaboration, in prep.}. Briefly, flux-calibrated spectra were extracted with pipelines developed by the ESA NIRSpec Science Operations Team and the NIRSpec GTO team, generally adopting the same algorithms as the STScI pipeline. Path-loss corrections were performed assuming point-like sources located at the position of the target object within the micro-shutter. In the DEEP tier, the one-dimensional PRISM spectra extracted over a shutter-size aperture reach a continuum sensitivity ($3 \sigma$) of $\sim 6$-$40 \times 10^{-22} \, \mathrm{erg \, s^{-1} \, cm^{-2} \, \AA^{-1}}$ (apparent magnitude of $m \sim 27.2$-$29.1$) per spectral pixel at $\sim 2 \, \mathrm{\upmu m}$ \citep{2023Natur.621..267W}.

\subsubsection{Data analysis}
\label{sssec:Data_analysis}

Having performed an automated spectral fitting routine with \program{bagpipes} \citep[Bayesian Analysis of Galaxies for Physical Inference and Parameter EStimation; ][]{2018MNRAS.480.4379C} on the PRISM spectra within the DEEP tier, spectroscopic redshift estimates were confirmed by visual inspection independently by at least two team members. The results of consolidated spectroscopic redshift confirmation will be described in a forthcoming paper \citetext{Bunker et al. 2023c, in prep.; see also \citealt{2023A&A...677A..88B}}. Using the inferred redshifts as a strict prior, the NIRSpec PRISM and R1000 spectra were fitted independently using the \program{ppxf} software \citep{2017MNRAS.466..798C} to obtain emission-line fluxes and accurate redshifts \citep[e.g.][]{2023arXiv230408516C}.

\lya\ emission in reionisation-era galaxies ($z > 5.8$) was identified during visual inspection, resulting in a sample of eight galaxies in the DEEP tier with secure detections in both the PRISM and R1000 spectra, and one additional galaxy where \lya\ is only detected in the PRISM. We complement our sample with seven LAEs identified in one of the MEDIUM tiers \citep{2023arXiv230602471J}.\footnote{We note \citet{2023arXiv230602471J} report four additional LAEs in which the signal-to-noise ratio (SNR) was insufficient to be included in the current sample.} This sample notably includes the extreme $z = 7.276$ LAE JADES-GS-z7-LA \citep[ID 10013682;][]{2023A&A...678A..68S}, while the intrinsically UV-brightest source in our sample (ID 12637 at $z = 7.66$) was previously identified as the LAE z7-GSD-3811 by \citet{2016ApJ...826..113S}.

The spectral properties, including those of \lya, were obtained using a custom fitting routine of the PRISM and R1000 spectra. Specifically, these properties include the \lya\ flux (measured from the R1000 spectra) and EW (measured in the PRISM spectra), the velocity offset ($\Delta v_\text{\lya}$; measured in the R1000 spectra), and finally the escape fraction ($f_\text{esc, \lya}$). The details of these measurements, and related empirical diagnostics derived from the measured line fluxes, are described in a companion paper \citep{2023arXiv230604536S}. Briefly, systemic redshifts were established from a SNR-weighted combination of Gaussian functions individually fitted to detected strong rest-frame optical emission lines in the R1000 spectra, specifically the $\OII \, \lambda \, 3727, 3730 \, \mathrm{\AA}$ doublet (\OII\ hereafter), \Hbeta, the $\OIII \, \lambda \, 4960, 5008 \, \mathrm{\AA}$ lines (\OIII), and \Halpha. Integrated line fluxes of these strong lines were determined from the fitted Gaussian profiles, taking into account the continuum level as measured in adjacent spectral regions; we turned to the PRISM spectra for lines that remained undetected in the R1000 spectra. The \lya\ flux was similarly determined by fitting a Gaussian profile to the R1000 measurements, which benefit from a higher sensitivity for emission lines.

We considered different methods of determining the \lya\ velocity offset; however, given the asymmetry of several \lya\ spectral profiles, which introduced systematic offsets between the peak of the best-fit Gaussian profile and the observed line profile, we opted to measure the velocity offset from the centre of the observed peak pixel of the line. The uncertainty of this measurement is conservatively estimated to be the width of a single spectral pixel, translating to $\ssim 100 \, \mathrm{km \, s^{-1}}$. \lya\ escape fractions are inferred by dividing the observed, dust-corrected \lya/\Halpha\ ratio (or \lya/\Hbeta\ if \Halpha\ is not available) by the intrinsic luminosity ratio of $L_\text{\lya}/L_\Halpha = 8.2$ under Case-B recombination, $n_e = 100 \, \mathrm{cm^{-3}}$, and $T_e = \num{10000} \, \mathrm{K}$ \citep{1989agna.book.....O}. Here, we report the main properties of the $17$ LAEs in \cref{tab:LAE_properties}.

Spectroscopically confirmed NIRSpec targets were matched to NIRCam photometric candidates \citep{2023ApJS..269...16R} where possible, noting that several sources fall outside the NIRCam footprint (including the LAE IDs 12637, 15362, and 10056849). For consistency with the samples from the FRESCO and MUSE surveys (\cref{ssec:MUSE_observations,ssec:FRESCO_observations}), we adopted absolute UV magnitudes $M_\text{UV}$ derived from NIRCam photometry probing a rest-frame wavelength of $\sim 1500 \, \mathrm{\AA}$ (F115W at $5.7 < z \lesssim 7.2$ or F150W at $7.2 \lesssim z < 10$) for all sources. We measured UV slopes ($\beta_\text{UV}$) of all LAEs from the PRISM spectra \citep[see][]{2023arXiv230604536S}, verifying that their normalisation at $1500 \, \mathrm{\AA}$ shows good agreement ($\ssim 20\%$) with the NIRCam UV magnitudes (adopting the NIRSpec value if NIRCam data was not available). As detailed in \citet{2023Natur.621..267W}, UV slopes of galaxies other than the $17$ LAEs were measured directly from the PRISM spectra for the sample of galaxies with median SNR higher than $3$ on the continuum (per spectral pixel, whose sizes are chosen adaptively such that the full width at half maximum (FWHM) is covered by 3.5 pixels; e.g. \citealt{2023arXiv230602467B}); otherwise, they were derived from the NIRCam photometry.

\begin{figure}
	\centering
	\includegraphics[width=\linewidth]{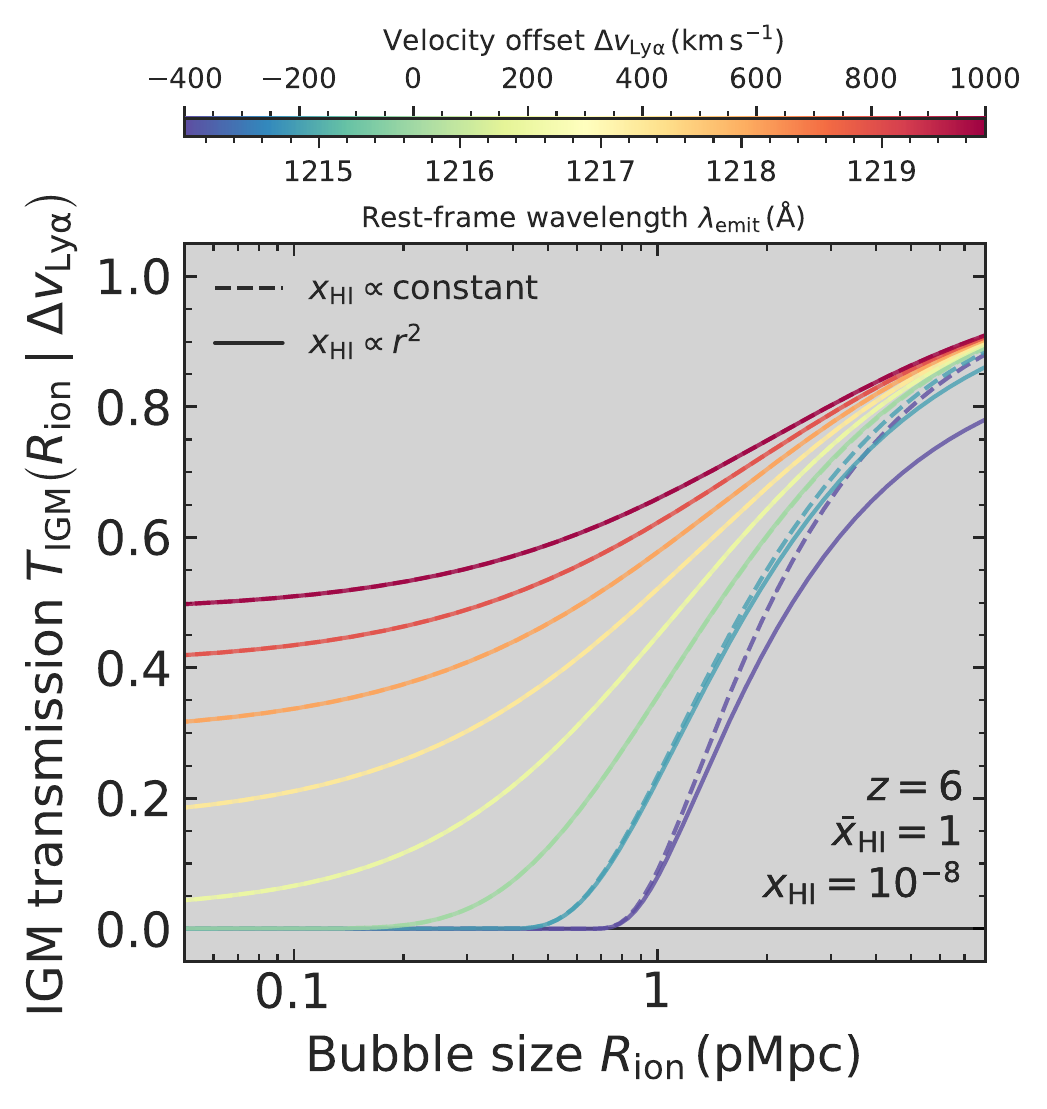}
	\caption{IGM transmission $T_\text{IGM}$ for a \lya\ emitting source at $z = 6$ for a given \lya\ velocity offset $\Delta v_\text{\lya}$ (coloured according to the scale shown on the top) as a function of the size of the ionised bubble, $R_\mathrm{ion}$. Results are shown both for a residual neutral fraction within the bubble scaling as $x_\text{\HI} \propto r^2$ (solid lines) and where it is constant with radius (dashed lines). Transmission curves in the top panel assume a global neutral hydrogen fraction of $\bar{x}_\text{\HI} = 0.01$ appropriate for the later stages of reionisation, whereas the bottom panel shows the transmission in a patchy reionisation scenario, where photons encounter gas that is still fully neutral upon exiting the ionised bubble ($\bar{x}_\text{\HI} = 1$).
	}
	\label{fig:Transmission_bubble_sizes}
\end{figure}
\begin{table*}[h]
    \centering
    \footnotesize
    \caption{Inferred minimum ionised bubble sizes.}
    \begin{tabular}{lll|lll|ll}
        \multicolumn{3}{c}{} & \multicolumn{3}{c}{Fixed $\bar{x}_\text{\HI} = 1$} & \multicolumn{2}{c}{Evolving $\bar{x}_\text{\HI}(z)$} \\
        \midrule
        ID & JADES source name & $z_\text{spec}$ & $R_\text{ion}^\text{req}$ & $R_\text{ion, LAE}^\text{pred}$ & $R_\text{ion, tot}^\text{pred}$ & $R_\text{ion}^\text{req}$ & $R_\text{ion, tot}^\text{pred}$ \\
        & & & $(\mathrm{pMpc})$ & $(\mathrm{pMpc})$ & $(\mathrm{pMpc})$ & $(\mathrm{pMpc})$ & $(\mathrm{pMpc})$ \\
        \midrule
        \csvreader[late after line=\\, head to column names]{Bubble_sizes.csv}{}{\NID & \ID & \zspec & \Rion & \RionLAE & \Riontot & \RiongxHI & \RiontotgxHI}
        \bottomrule
    \end{tabular}
    \tablefoot{
        Listed properties of each LAE are its NIRSpec ID, full JADES identifier, spectroscopic redshift ($z_\text{spec}$), and inferred minimum ionised bubble sizes ($R_\text{ion}^\text{req}$), if at all required to explain the observed velocity offset (\cref{tab:LAE_properties}). Estimates are reported both assuming the neutral hydrogen fraction outside the bubble is fixed ($\bar{x}_\text{\HI} = 1$) or evolving with redshift ($\bar{x}_\text{\HI}(z)$; see \cref{ssec:Ionised_bubble_sizes} for details). Additional columns for both scenarios list the bubble sizes that can be produced by the LAE on its own ($R_\text{ion, LAE}^\text{pred}$) or by all sources contained within the bubble ($R_\text{ion, tot}^\text{pred}$), assuming a fiducial age of $t_* = 50 \, \mathrm{Myr}$ and $f_\text{esc, LyC} = 5\%$ (see \cref{ssec:Ionised_bubble_growth} for details).
    }
    \label{tab:Bubble_sizes}
\end{table*}

\subsection{MUSE HUDF surveys}
\label{ssec:MUSE_observations}

Since the JADES coverage of GOODS-S fully incorporates the HUDF, we further supplement our sample of spectroscopically confirmed galaxies with sources identified in the publicly available MUSE HUDF surveys \citep{2017A&A...608A...1B}. From the DR2 catalogue presented in \citet{2023A&A...670A...4B}, we selected galaxies with a confident spectroscopic redshift (\textsc{zconf} of either $2$ or $3$; e.g. \citealt{2023MNRAS.523.5468S}) at $z > 5.5$, which are all spectroscopically confirmed solely via \lya\ as a result of the wavelength range covered by MUSE \citep[e.g.][]{2017A&A...608A...1B}. Since this prohibits us from obtaining a \lya\ velocity offset, we do not include these LAEs in our main sample, only considering them in the environmental analysis of other LAEs instead (\cref{ssec:Ionised_bubble_growth}). As for the JADES galaxies, we measured UV magnitudes and UV slopes based on the NIRCam photometric catalogue (\cref{sssec:Data_analysis}).

Having performed a cross-matching procedure, however, we find that all five JADES LAEs observable with MUSE (i.e. within its footprint and at $z \lesssim 6.7$) are contained as robustly confirmed galaxies ($\text{\textsc{zconf}} \geq 2$) in the MUSE HUDF DR2 catalogue. Specifically, IDs 19606, 9365, 19342, 14123, and 16625 in \cref{tab:LAE_properties} are matched to the respective MUSE IDs 3203, 3089, 547, 6231, and 852 in \citet{2023A&A...670A...4B}; these sources were removed from the MUSE sample to avoid double counting. One of these sources is the $z = 5.889$ LAE (ID 19606) where \lya\ emission is only seen in the PRISM spectrum; in this case, we adopt the offset of the \lya\ line observed by MUSE from the systemic redshift measured by NIRSpec, $\Delta v \approx 50 \, \mathrm{km \, s^{-1}}$, as the \lya\ velocity offset. In the other four cases, the \lya\ offset as observed by NIRSpec is further redshifted by between $50 \, \mathrm{km \, s^{-1}}$ and $200 \, \mathrm{km \, s^{-1}}$ compared to what was reported in \citet{2023A&A...670A...4B}. While unlikely to be affected by the underlying continuum which remains undetected in the medium-resolution R1000 spectra \citep{2023arXiv230604536S}, this systematic difference may be due to potential calibration issues. Alternatively, it could suggest the \lya\ emission is spatially extended with the two instruments probing different spatial regions with mildly varying peak wavelength as a result of the spatial variation in the column density of neutral hydrogen \citep[e.g.][]{2020A&A...635A..82L}.

\subsection{FRESCO}
\label{ssec:FRESCO_observations}

In addition to the JADES imaging and spectroscopy, we made use of NIRCam wide-field slitless spectroscopy (WFSS) taken over GOODS-S as part of FRESCO \citep[\textit{JWST} programme 1895, PI: Oesch;][]{2023MNRAS.525.2864O}. For the public FRESCO data, we follow the NIRCam WFSS data reduction routine outlined in \citet{2023ApJ...953...53S} and emission-line identification algorithms presented in \citet{2023arXiv230210217H}. After a careful visual inspection of extracted slitless spectra of galaxies with NIRCam photometric redshifts of $z > 5$, we spectroscopically confirmed hundreds of galaxies within the JADES footprint via \Halpha\ emission at $4.9 < z < 6.5$ or $\OIII \, \lambda \, 5008 \, \mathrm{\AA}$ emission at $6.8 < z < 8.9$ \citetext{Sun et al., in prep.}. A future work will present detailed descriptions of the WFSS data reduction and catalogues of these spectroscopically confirmed galaxies. Cross matching the MUSE and FRESCO samples resulted in ten LAEs originally identified in the MUSE HUDF surveys for which a new systemic redshift is found in FRESCO, resulting in implied \lya\ velocity offsets ranging from $\Delta v_\text{\lya} \sim 100 \, \mathrm{km \, s^{-1}}$ to $400 \, \mathrm{km \, s^{-1}}$ (we note that NIRCam WFSS redshifts may be subject to a calibration uncertainty of $40$ to $80 \, \mathrm{km s^{-1}}$). We removed these objects from the FRESCO sample and updated the systemic redshift of the corresponding sources in the MUSE sample. Finally, we confirm the spectroscopic redshift $z = 7.955$ of GSDY-2209651370, which was previously identified as an LAE by \citet{2023ApJ...948...54R}. The systemic redshift found in FRESCO implies a \lya\ velocity offset of $\Delta v_\text{\lya} \approx 230 \, \mathrm{km \, s^{-1}}$.

\subsection{Combined spectroscopically confirmed galaxy samples}
\label{ssec:Combined_spectroscopically_confirmed_galaxy_samples}

In summary, our main sample comprises $17$ LAEs at $5.8 \lesssim z \lesssim 8$ observed as part of JADES, five of which have previously been identified as part of the MUSE HUDF surveys. To study their environments, we further consider $88$ LAEs at $5.5 \lesssim z \lesssim 6.7$ from the MUSE HUDF surveys, as well as spectroscopically confirmed galaxies (without confirmed \lya\ emission) from JADES and FRESCO data, totalling respectively $22$ and $243$ sources in the same redshift range.

\begin{figure*}
	\centering
	\includegraphics[width=\linewidth]{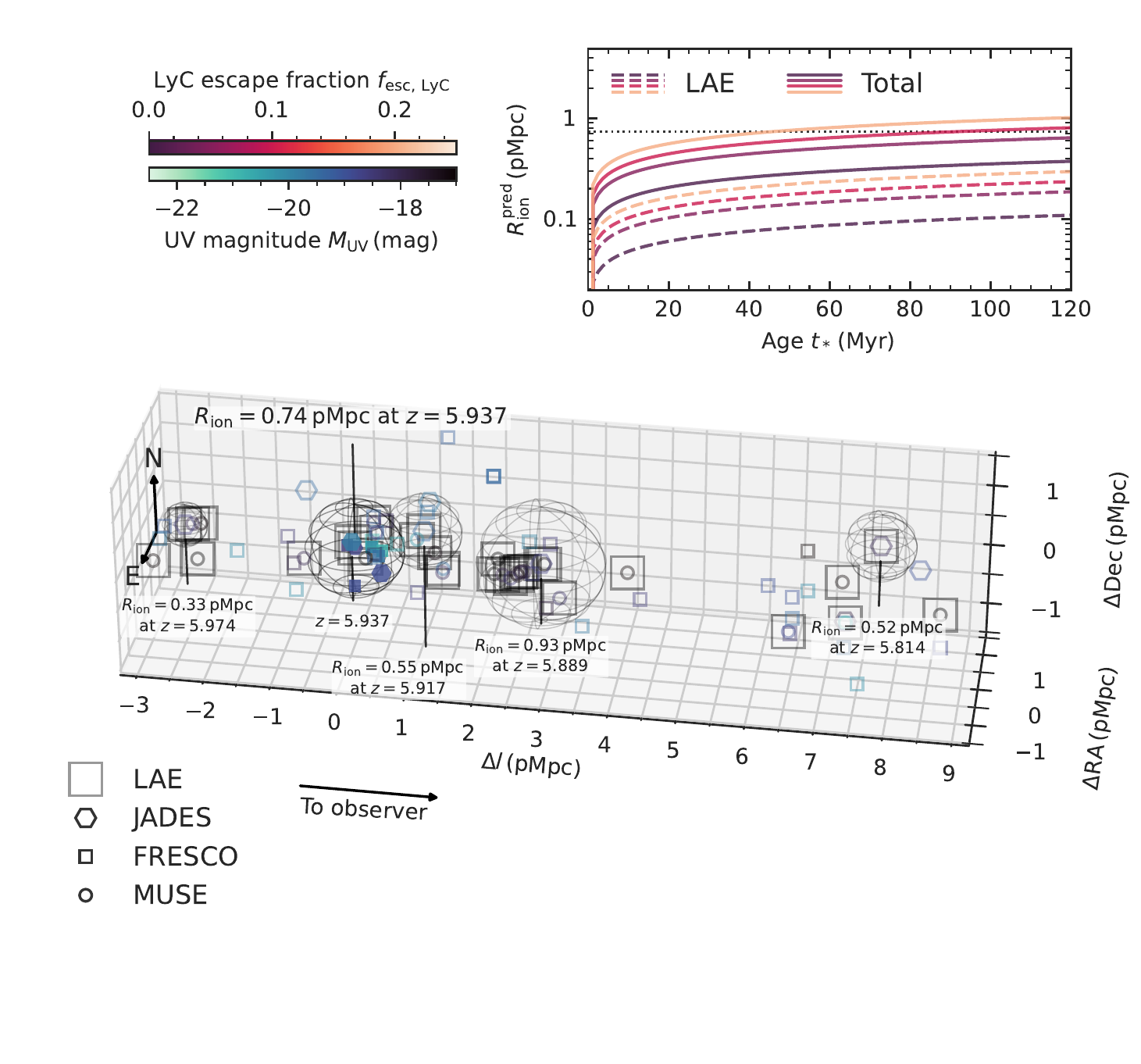}
	\caption{Three-dimensional visualisation of the complex of spectroscopically confirmed $z \sim 5.9$ galaxies situated along the line of sight. JADES sources are shown by hexagons, FRESCO sources by squares, and MUSE sources by circles. Many of the objects in JADES and MUSE are observed to be LAEs (indicated by large grey squares). For the six LAEs observed in JADES, ionised bubbles whose sizes were inferred under the assumption that $\bar{x}_\text{\HI} = 1$ (see \cref{ssec:Ionised_bubble_sizes}) are shown by wireframe spheres centred on each LAE (see also annotations). Galaxies contained within the ionised bubble of the main $z = 5.937$ LAE considered here (ID 6002) are indicated with filled markers (open markers otherwise); all sources are furthermore coloured according to their absolute UV magnitude $M_\text{UV}$. The inset in the top right corner illustrates the bubble size calculation discussed in \cref{ssec:Ionised_bubble_growth} \citep[see also][]{2023A&A...678A..68S}. It shows the predicted bubble size $R_\text{ion}^\text{pred}$ for ID 6002 as a function of time since an ionising source has turned on, $t_*$, for a range of LyC escape fractions ($1\%$, $5\%$, $10\%$, and $20\%$), either for the LAE itself (dashed lines) or all sources within its required bubble size. This simple calculation indicates that the neutral hydrogen fraction outside the ionised bubble is likely $\bar{x}_\text{\HI} \ll 1$, as expected at $z \sim 5.9$ (\cref{ssec:LAEs_and_their_environments}).
	}
	\label{fig:LAE_complex_z59}
\end{figure*}

\section{Results and discussion}
\label{sec:Results_and_discussion}

\subsection{Ionised bubble sizes}
\label{ssec:Ionised_bubble_sizes}

In a patchy reionisation scenario, where a galaxy embedded in a neutral IGM emits \lya\ at a velocity offset\footnote{It is expected that as a result of complex radiative transfer in the interstellar and circumgalactic medium, the \lya\ emission of a galaxy (before IGM processing) typically has a line profile dominated by a redshifted peak with respect to its rest frame \citep{2021MNRAS.501.5757M, 2023MNRAS.523.3749B}, as seen in observations \citep[e.g.][]{2021ApJ...908...36H, 2021MNRAS.508.1686W}.} of $\Delta v_\text{\lya} \lesssim 300 \, \mathrm{km \, s^{-1}}$ (as is typical for our sample; \cref{tab:LAE_properties}), the photons experience substantial absorption unless the galaxy is situated in a highly ionised region with radius of the order of $R_\text{ion} \gtrsim 0.1 \, \text{pMpc}$ \citep[e.g.][]{2020MNRAS.499.1395M, 2023arXiv230600487U}. Transmission of \lya\ photons in a fully neutral IGM, on the other hand, is possible even without such an ionised bubble, as long as they are sufficiently redshifted \citep[cf.][]{2023A&A...677A..88B}. This is illustrated in \cref{fig:Transmission_bubble_sizes}, which shows the IGM transmission as a function of the ionised bubble size, following the modelling prescription in \citet{2020MNRAS.499.1395M}. Specifically, the transmission was calculated using a two-zone model, first considering the trajectory of a photon through an ionised bubble\footnote{For simplicity, we work under the assumption that radiation from a central ionising source is emitted isotropically, resulting in a spherical geometry of ionised bubbles. We will further discuss the implications of this assumption in \cref{ssec:LAEs_and_their_environments}.} of size $R_\mathrm{ion}$ before encountering the neutral IGM \citep[see also][]{2000ApJ...542L..75C, 2004ApJ...613...23M}. Following \citet{2020MNRAS.499.1395M}, the gas in the ionised bubble was assumed to have a temperature of $T = 10^4 \, \mathrm{K}$ while the neutral IGM has $T = 1 \, \mathrm{K}$; both are considered to have mean cosmic density and be at rest with respect to the central source. We assumed the bubble to be highly ionised, where the residual neutral fraction is either constant with radius (fixed at $x_\text{\HI} = 10^{-8}$), corresponding to homogeneous reionisation of the bubble, or scales as $x_\text{\HI} \propto r^2$ with normalisation $x_\text{\HI} (r = 0.1 \, \mathrm{pMpc}) = 10^{-8}$, as would be expected with a central source of ionisation. The implications of these assumptions will be discussed further below.

Here, we aim to infer a first-order estimate in the form of a lower limit on the size of the ionised region that allows each observed \lya\ line in our sample to be observed close to the systemic redshift. We note that this framework is expected to become inaccurate towards the end of reionisation, where ionised bubbles start to overlap and the neutral fraction of the IGM uniformly approaches $\bar{x}_\text{\HI} \sim 0$ \citep{2023arXiv230411192L}, such that the distinction between ionised bubbles and neutral IGM breaks down. This is illustrated in the top panel of \cref{fig:Transmission_bubble_sizes}, showing that in our model (assuming a static IGM) even without a large ionised region, a low neutral fraction of the IGM accommodates significant transmission of redshifted \lya\ ($\Delta v_\text{\lya} \gtrsim 0$). We note, however, that a bulk infalling motion of the IGM lowers such transmission: in the rest frame of the IGM, having a peculiar velocity directed towards the central emitter, light will appear blueshifted thus `resetting' any redshift the \lya\ photons may have. This effectively shrinks the size of a potential ionised bubble, requiring a larger physical size for \lya\ to escape. Patches of residual neutral hydrogen within ionised bubbles, as predicted by cosmological hydrodynamical simulations of the reionisation process, would similarly cause an additional suppression of transmitted \lya\ photons on the red side ($\Delta v_\text{\lya} > 0$), though minimally as this mainly affects transmission on the blue side \citep[$\Delta v_\text{\lya} \lesssim 0$;][]{2023arXiv230805800K}. Instead, our approach is designed to give an insight into the \emph{minimum} sizes of local ionised regions of the highest-redshift ($z \gtrsim 6$) LAEs while being agnostic to the evolution of the global neutral fraction, which instead will be discussed in a different work \citep{2023arXiv230602471J}.

For this reason, we calculated ionised bubble sizes under two opposite extremes. The first method simply assumes the IGM (i.e. all gas outside an ionised bubble), has a residual \emph{global} neutral hydrogen fraction of $\bar{x}_\text{\HI} = 1$ \citep[as in][]{2020MNRAS.499.1395M}. The second method instead assumes the global neutral fraction follows a smooth redshift evolution, $\bar{x}_\text{\HI} (z)$. This scenario of a homogeneously ionised Universe is likely appropriate for the end stages of reionisation and afterwards \citep[$z < 6$; e.g.][]{2022MNRAS.514...55B}. Noting recent evidence points towards a (very) late cosmic reionisation history where this global neutral fraction could still be non-negligible at the redshift range considered here \citep[e.g. $\bar{x}_\text{\HI} \sim 5\%$ at $z \sim 5.6$;][]{2022ApJ...932...76Z}, we adopted the rapid and late reionisation presented as `Model II' in \citet{2020ApJ...892..109N} as a fiducial model for the evolution of $\bar{x}_\text{\HI} (z)$.

\begin{figure*}
	\centering
	\includegraphics[width=\linewidth]{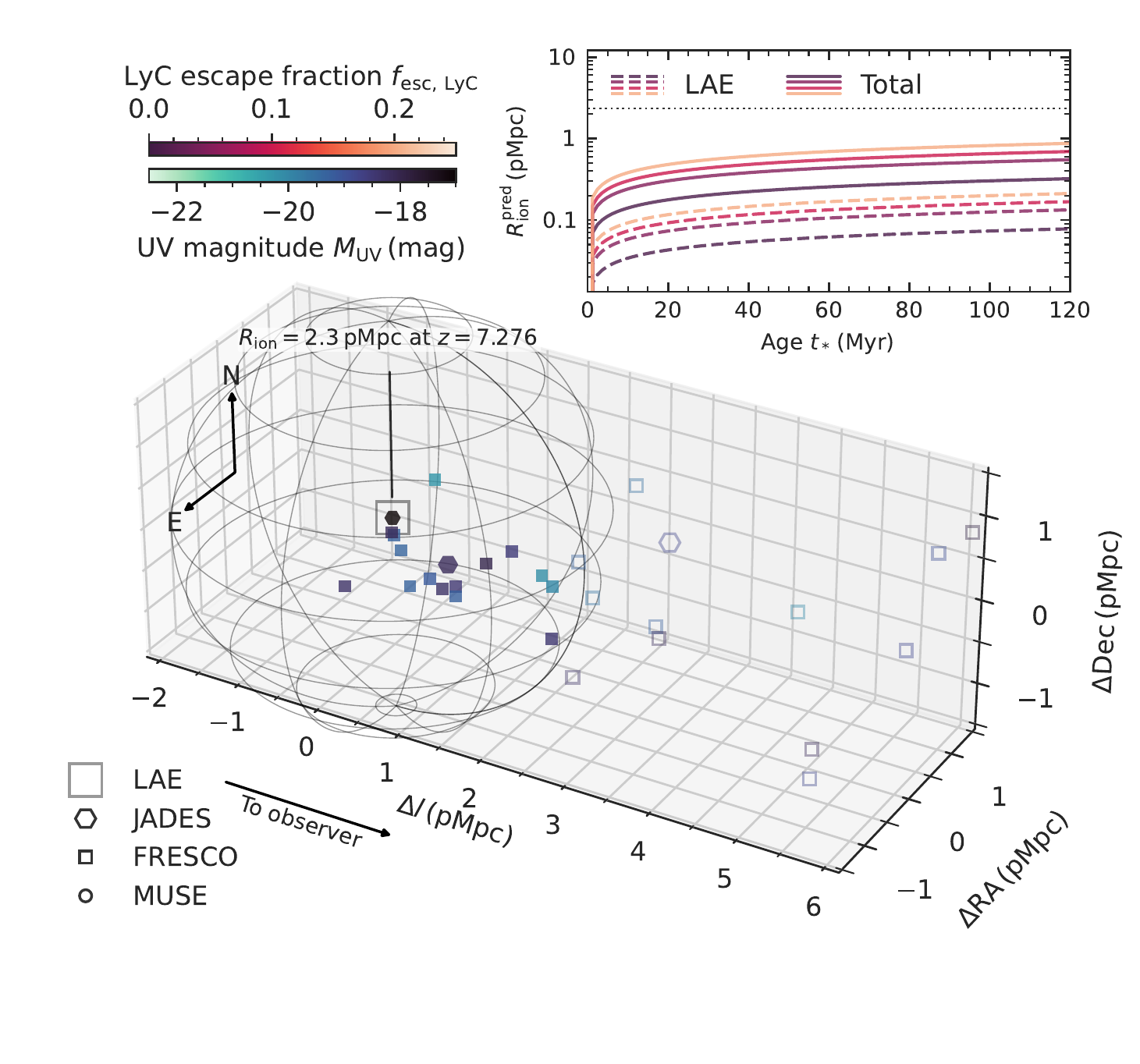}
	\caption{Similar to \cref{fig:LAE_complex_z59}, but for spectroscopically confirmed galaxies around JADES-GS-z7-LA, the extreme $z = 7.276$ LAE \citep{2023A&A...678A..68S}.
	}
	\label{fig:LAE_complex_z73}
\end{figure*}

Using this model of IGM absorption, we estimate how large the ionised region needs to be in the conservative case where all \lya\ escapes the galaxy in the first place: that is, we infer the minimum required bubble size $R_\text{ion}^\text{req}$ for which
\begin{equation}
    \label{eq:Bubble_size_criterion}
    T_\text{IGM} \left( \left. R_\text{ion}^\text{req} \, \right| \Delta v_\text{\lya} \right) = f_\text{esc, \lya} \, ,
\end{equation}
\noindent where $f_\text{esc, \lya}$ is inferred from the \lya/\Halpha\ ratio and represents the observed fraction of \lya\ photons intrinsically produced in \HII\ regions \citep{2023arXiv230604536S}. We thus approximate all \lya\ flux has the same velocity whereas in reality, the line is broadened. We expect this approximation to be reasonably accurate, however, as additional absorption of photons bluewards of the peak wavelength is balanced by enhanced transmission of their red counterparts, particularly when the intrinsic line profile (i.e. before, or in the absence of, IGM attenuation) has a dominant, asymmetric redshifted peak (as seen at lower redshift; \cref{app:Line_profile}). We note that current observations prevent an accurate measurement of the intrinsic line width: our observed line profiles are only marginally resolved at a resolution of $R \sim 1000$, typically spanning only a few spectral channels in the first place \citep[see][]{2023arXiv230604536S}, while the exact spectral resolution of the NIRSpec micro-shutters is dependent on the (unknown) morphology of the \lya\ emission \citep{2023arXiv230809742D}.

This method effectively provides a lower limit to the ionised bubble size, since the number of \lya\ photons that travel through the IGM towards our line of sight is further lowered by any additional absorption or loss of photons that can be attributed to (i) dust within the interstellar medium, (ii) an IGM infall velocity (discussed above), (iii) a higher residual neutral fraction within the bubble, or (iv) resonant scattering on the scales of the circumgalactic medium (CGM) that extend outside the NIRSpec micro-shutter. Each of these effects would require an even larger ionised region to reconcile the observed \lya\ flux with the intrinsic \lya\ production via recombinations probed by the \Halpha\ flux, the two components that enter into the \lya\ escape fraction estimated via the \lya/\Halpha\ ratio. We note, however, the impact of dust in the sample of LAEs considered here is expected to be minimal, as measured by the effectively non-existent Balmer decrement (consistent with $E(B-V) = 0$ within uncertainties) indicating that nebular dust attenuation in these galaxies is negligible \citep{2023arXiv230604536S, 2023arXiv230603931S}. Moreover, while \lya\ emission around star-forming galaxies at $2 \lesssim z \lesssim 6$ has been shown to extend to scales of $\ssim 5 \, \mathrm{kpc}$ \citep[e.g.][]{2008ApJ...681..856R, 2020A&A...635A..82L} which roughly corresponds to the extent of a NIRSpec MSA shutter at $z \sim 7$ \citep{2022A&A...661A..81F}, it is unclear whether this \lya\ emission can actually be directly tied to the central galaxy. Indeed, there are indications that such extended emission may be largely produced in situ \citep[e.g. by collisional excitation or faint, unresolved neighbouring LAEs;][]{2021A&A...650A..98W, 2021A&A...647A.107B}, while the central component is expected to predominantly originate in recombinations traced by Balmer emission \citep[e.g.][]{2021MNRAS.506.5129B, 2023MNRAS.522.4275R}.

We opted for the quadratic radial profile of the residual neutral fraction within the ionised bubble (i.e. $x_\text{\HI} \propto r^2$), noting the results do not strongly depend on this assumption for all but the most blueshifted \lya\ emission (cf. \cref{fig:Transmission_bubble_sizes}). The resulting ionised bubble sizes, if required, are reported in \cref{tab:Bubble_sizes} for both assumptions on the global neutral hydrogen fraction discussed above. In case \cref{eq:Bubble_size_criterion} has no solution for all $R_\text{ion} > 10^{-3} \, \mathrm{pMpc}$ (which is the case for most $z < 6.5$ LAEs in the evolving $\bar{x}_\text{\HI}$ scenario), we conclude a bubble does not need to be invoked for the model to reconcile the observations: in this case, the minimum IGM transmission (i.e. that without a bubble) is already higher than the \lya\ escape fraction, indicating that given the modelled IGM transmission our conservative assumption on the IGM accounting for all absorption is no longer valid. The sizes inferred in the case where $\bar{x}_\text{\HI} = 1$ are furthermore shown in \cref{fig:LAE_on-sky_distribution}. The conservatively estimated uncertainty on the measured \lya\ velocity offsets of $\ssim 100 \, \mathrm{km \, s^{-1}}$ (\cref{sssec:Data_analysis}) translates to a $\ssim 0.1$-$0.15 \, \mathrm{pMpc}$ uncertainty on the required sizes, significantly smaller than the estimated bubble sizes in the majority of cases.

In a patchy reionisation scenario where all LAEs are still surrounded by a fully neutral IGM (i.e. under the assumption that $\bar{x}_\text{\HI} = 1$), we find ionised bubble sizes of the order of $R_\text{ion}^\text{req} \sim 0.1$-$1 \, \mathrm{pMpc}$, except for the extreme $z = 7.276$ LAE JADES-GS-z7-LA (ID 10013682), for which we find $R_\text{ion} \approx 2.4 \, \mathrm{pMpc}$ \citep[in agreement with][]{2023A&A...678A..68S}. The largest ionised bubbles are generally inferred for the highest equivalent width LAEs: for instance, the two largest bubble sizes after JADES-GS-z7-LA are those around the $z \sim 5.889$ and $z \sim 6.204$ LAEs, both having $\text{EW}_\text{\lya} \sim 90 \, \mathrm{\AA}$. Given our sample does not have a large variation in \lya\ velocity offsets, this can be explained by the strong correlation between $\text{EW}_\text{\lya}$ and \lya\ escape fraction \citep{2023arXiv230604536S} that can be interpreted as the LAEs being characterised by a similar $\xi_\text{ion}$ (implying a similar ratio between the strength of \Halpha\ and the UV continuum) such that the ratio of \lya\ to the UV continuum (i.e. the \lya\ EW) tightly follows the ratio between \lya\ and \Halpha. In the case where the global neutral hydrogen fraction evolves with redshift ($\bar{x}_\text{\HI}(z)$), the \lya\ transmission of many of the lower-redshift LAEs is found to be consistent with not having a bubble, while for sources at $z \gtrsim 6.5$ we still find minimum bubble sizes of $R_\text{ion}^\text{req} \sim 0.1$-$0.5 \, \mathrm{pMpc}$.

\subsection{Ionised bubble growth}
\label{ssec:Ionised_bubble_growth}

The expected connection between LAEs and ionised bubbles is corroborated by the expectation that the physical conditions necessary for an efficient production and escape of Lyman-continuum (LyC) photons required to create these bubbles -- at least if the LyC escape fraction is less than unity -- should be accompanied by intrinsically (i.e. before IGM attenuation) luminous \lya\ lines, both from a theoretical perspective as well as empirical evidence at $z \sim 2$ \citep{2022MNRAS.512.5960M, 2022MNRAS.510.4582N}. On the other hand, galaxies observed to have bright \lya\ emission do not necessarily have to be efficient LyC leakers themselves \citep{2023arXiv230408526C}. This degeneracy can be broken by measuring the sizes of ionised regions required to explain the observed \lya\ properties and testing whether these agree with their predicted ionised output of the LAE and any potential neighbouring galaxies. The immediate environments of reionisation-era LAEs inside ionised bubbles therefore provide a unique laboratory in which we can start to establish which sources are the dominant ionising agents.

We therefore proceed by obtaining first-order estimates for the sizes of local, ionised bubbles that are expected to be in place for our sample of LAEs. Following \citet{2023A&A...678A..68S}, we calculate the bubble growth as a function of time $t_*$ since a central source started producing LyC photons using a simplified solution of the equation describing ionisation-front propagation with a central ionising source \citep[cf.][]{2000ApJ...542L..75C}. We assumed this source has a given production rate of ionising photons $\dot{N}_\text{ion}$ and escape fraction $f_\text{esc, LyC}$. Neglecting the accelerated bubble expansion due to the Hubble flow and under the assumption of a subdominant recombination rate, which at redshift $z \lesssim 8$ is a good approximation,\footnote{Within this framework of a central ionising source at $z = 6$ ($z = 8$), the recombination rate within its ionised bubble (the rightmost term of equation (3) in \citealt{2000ApJ...542L..75C}) becomes comparable to the ionisation rate only when $t_* \gtrsim 600 \, \mathrm{Myr}$ ($300 \, \mathrm{Myr}$) assuming a clumping factor of $C_\text{\HII} = 3$ \citep[e.g.][]{2009MNRAS.394.1812P}.} the size evolution is given by equation (6) in \citet{2020MNRAS.499.1395M},
\begin{equation}
    \label{eq:Bubble_growth}
    R_\text{ion} (t_*) \approx \left( \frac{3 f_\text{esc, LyC} \, \dot{N}_\text{ion} \, t_*}{4 \pi \, \bar{n}_\text{H} (z)} \right)^{1/3} \, ,
\end{equation}

\noindent where $\bar{n}_\text{H} (z)$ is the mean hydrogen number density at redshift $z$.

Assuming star formation as the dominant source of ionising photons, bubble sizes that can be produced by the LAE on its own ($R_\text{ion, LAE}^\text{pred}$) or by the LAE and its direct neighbours ($R_\text{ion, tot}^\text{pred}$) are reported in \cref{tab:Bubble_sizes}. For the latter calculation, we simply used a combined ionising photon production rate for all spectroscopically confirmed sources contained within the bubble whose minimum size required for \lya\ to escape was inferred in \cref{ssec:Ionised_bubble_sizes} (i.e. $\dot{N}_\text{ion, tot} = \sum \dot{N}_\text{ion, i}$ with $i$ iterating over each neighbour). In calculating $\dot{N}_\text{ion}$ for the LAEs, we converted the UV magnitude $M_\text{UV}$ using the ionising photon production efficiency $\xi_\text{ion}$ that was directly measured (with median $\xi_\text{ion} \approx 10^{25.59} \, \mathrm{Hz \, erg^{-1}}$; \citealt{2023arXiv230604536S}). For neighbouring sources, lacking a direct measurement of $\xi_\text{ion}$, we took their UV magnitude $M_\text{UV}$ and UV slope $\beta_\text{UV}$ to estimate $\dot{N}_\text{ion}$ under the assumption that the ionising spectrum assumes a double power-law shape, as described by equations (7) to (9) in \citet{2020MNRAS.499.1395M}. We assumed an ionising-continuum slope $\alpha = 2$ \citep{2023A&A...678A..68S}, noting that for a typical $\beta_\text{UV} = -2$ this corresponds to assuming $\xi_\text{ion} \approx 10^{25.58} \, \mathrm{Hz \, erg^{-1}}$ for the ionising photon production efficiency, comparable to what is directly measured for LAEs (e.g. \citealt{2023arXiv230604536S}; see also \citealt{2023MNRAS.526.1657T}; \citealt{2023MNRAS.523.5468S}).

For these estimates, we assumed fiducial values of $f_\text{esc, LyC} = 5\%$ \citep[as should be appropriate for these LAEs;][]{2023arXiv230604536S} and an age of $t_* = 50 \, \mathrm{Myr}$, noting the inferred bubble size does not strongly depend on small deviations in LyC escape fraction and age. In this simple framework that only counts the total number of ionising photons, there is a degeneracy between the two parameters: for instance, $f_\text{esc, LyC} = 5\%$ and $t_* = 50 \, \mathrm{Myr}$ are equivalent to having a LyC escape fraction of $10\%$ ($50\%$) for $25 \, \mathrm{Myr}$ ($5 \, \mathrm{Myr}$). Finally, we note these parameters are fundamentally bounded: the ionising photon production efficiency is unlikely to exceed $\xi_\text{ion} = 10^{26} \, \mathrm{Hz \, erg^{-1}}$ \citep[e.g.][]{2023MNRAS.526.1657T, 2023MNRAS.525.2422S}, the escape fraction is limited to a maximum of $100\%$, and periods of constant star formation are generally not expected to significantly exceed $200 \, \mathrm{Myr}$ at this early epoch \citep{2018ApJ...868...92T, 2023MNRAS.519..157W}. We again consider both assumptions on the global neutral hydrogen fraction discussed in \cref{ssec:Ionised_bubble_sizes}, though this only impacts the number of neighbours and thus the inferred $R_\text{ion, tot}^\text{pred}$, since the LAE itself is always contained within its own bubble. The resulting predicted sizes of ionised bubbles are presented in \cref{tab:Bubble_sizes}; we will discuss how these compare with the minimum sizes required based on the \lya\ transmission in the next section, first briefly discussing the environments of the LAEs in our sample.

\subsection{Reionisation-era LAEs and their environments}
\label{ssec:LAEs_and_their_environments}

To illustrate the potential importance of environment in our sample of LAEs, we show examples of the three-dimensional distributions of spectroscopically confirmed galaxies in \cref{fig:LAE_complex_z59,fig:LAE_complex_z73}. In these cases, as also seen in \cref{fig:LAE_redshift_distribution}, LAEs coincide with a large number of galaxies identified spectroscopically in FRESCO data \citep{2023arXiv231104270H}, which for the $z \sim 7.3$ case is supported by photometric selection of galaxies \citep{2023arXiv230605295E}. We further note the association of LAEs shown in \cref{fig:LAE_complex_z59} is relatively close (within $\ssim 10 \, \mathrm{pMpc}$) to the $z \sim 5.78$ LAE overdensity already reported by \citet{2021A&A...647A.107B}; indeed, both the $z \sim 5.78$ and $z \sim 5.93$ associations coincide with several of the first LAEs found in GOODS-S \citep{2003MNRAS.342L..47B, 2004ApJ...604L..13S, 2007MNRAS.376..727S}. The FRESCO data additionally confirms the presence of more than ten galaxies in small redshift slices around $z \sim 5.78$, $z \sim 5.93$, and $z \sim 7.25$, rendering each of them significant overdensities compared to the average field. Details of the clustering properties of these galaxy associations are discussed in \citet{2023arXiv231104270H}. In addition, close to the $z \sim 5.93$ overdensity there is another densely packed group of $z \sim 5.89$ LAEs in a `filament', in which the IGM is likely highly ionised leading to enhanced transmission of \lya.

\begin{figure}
	\centering
	\includegraphics[width=\linewidth]{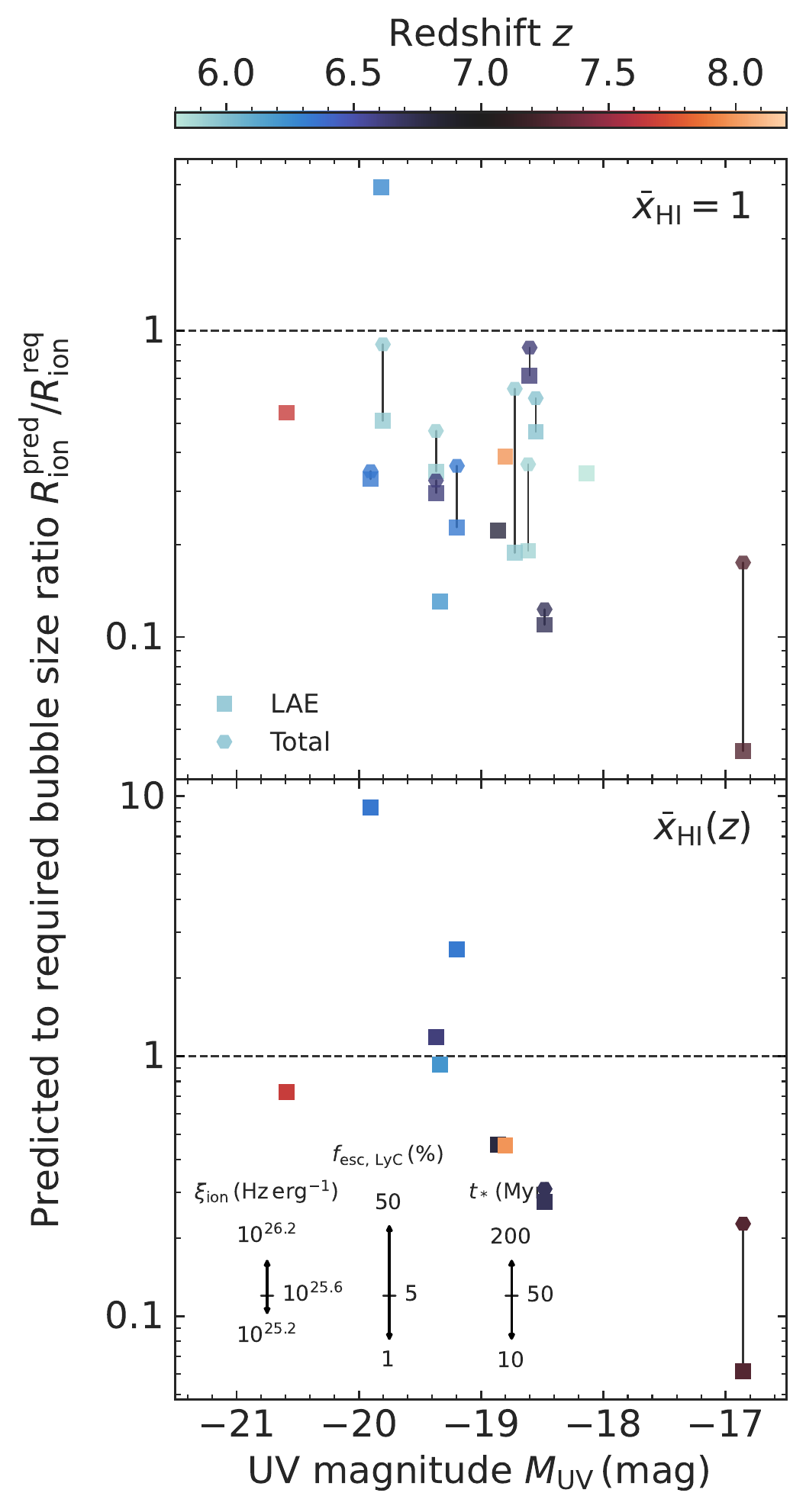}
	\caption{Comparison of the inferred bubble sizes ($R_\text{ion}^\text{req}$), if required to explain the observed velocity offset (\cref{tab:LAE_properties}), to predicted bubble sizes ($R_\text{ion, tot}^\text{pred}$; see \cref{ssec:Ionised_bubble_growth}) as a function of absolute UV magnitude $M_\text{UV}$ of the LAE. Inferred sizes are shown both assuming the neutral hydrogen fraction outside the bubble is fixed ($\bar{x}_\text{\HI} = 1$; top panel) or evolving with redshift ($\bar{x}_\text{\HI}(z)$, in which case a bubble around many of the lower-redshift sources is not found to be necessary; bottom panel). Predicted bubble sizes that can be produced by the LAE alone are shown by squares, while those with an additional ionising photon contribution by other sources contained within the bubble, if any, are indicated by hexagons. Predictions assume a fiducial age of $t_* = 50 \, \mathrm{Myr}$ and $f_\text{esc, LyC} = 5\%$; arrows in the bottom panel indicate how estimates shift under varying assumptions.}
	\label{fig:Bubble_size_comparison}
\end{figure}

For the main LAEs shown in \cref{fig:LAE_complex_z59,fig:LAE_complex_z73}, even when considering all detected sources located within the minimum required bubble size for \lya\ escape, the ionising photon budget appears to fall short of explaining the inferred bubble size (reconciliation requires all sources to leak LyC radiation at $f_\text{esc, LyC} = 20\%$ for $t_* \sim 50 \, \mathrm{Myr}$). In the case of the $z \sim 5.9$ LAE complex (\cref{fig:LAE_complex_z59}), this likely indicates that the neutral hydrogen fraction outside the ionised bubble is lower than the assumed $\bar{x}_\text{\HI} = 1$, at least for these overdense patches of the Universe: the fact that many LAEs are observed closely together (in particular the dense cluster around $z \sim 5.889$) shows that the ionised bubbles, required as a minimum around each LAE individually, together make up a significant volume fraction of (nearly) fully ionised hydrogen. This likely represents the stage of reionisation where individual bubbles have started to overlap and thus form larger ionised regions in which the simplified approach of isolated bubbles within a fully neutral IGM is no longer valid, as discussed in \cref{ssec:Ionised_bubble_sizes}. Indeed, independent methods indicate ionised bubble sizes of the order of $5 \, \mathrm{pMpc}$ at this redshift regime \citep[$5.7 < z < 6.1$;][]{2020MNRAS.494.1560M, 2023ApJ...950...66K}, in agreement with expectations from simulations which further show galaxy overdensities are more likely to reside in large bubbles \citep[e.g][]{2022MNRAS.510.3858Q, 2023arXiv230411192L}. While representing lower limits, our ionised bubble size estimates are considerably smaller than $R_\text{ion} \sim 100 \, \text{pMpc}$ recently inferred by \citet{2023arXiv230600487U} for brighter systems (typically $M_\text{UV} \lesssim -20 \, \mathrm{mag}$; note however that \citealt{2023arXiv230805800K} show the blue flux transmission at $z \sim 7$ implied by such large ionised regions is not measured in the \lya\ forest of quasars until $z \lesssim 3$). In the case with a smoothly evolving global neutral hydrogen fraction (where $\bar{x}_\text{\HI} (z = 6) = 0.01$; \citealt{2020ApJ...892..109N}) we infer ionised bubbles are not necessarily required to explain the observed \lya\ transmission properties for any of these sources (\cref{tab:Bubble_sizes}).

Conversely, the extreme $z = 7.276$ LAE (ID 10013682 or JADES-GS-z7-LA) shown in \cref{fig:LAE_complex_z73}, requires a large ionised bubble ($R_\text{ion}^\text{req} \sim 2 \, \mathrm{pMpc}$ even with a smoothly evolving global neutral fraction; \cref{tab:Bubble_sizes}), as discussed in \citet{2023A&A...678A..68S}. Such a bubble is difficult to attribute to the relatively UV-faint galaxies identified in its environment (the FRESCO data being able to identify sources with $M_\text{UV} \lesssim -18 \, \mathrm{mag}$; \cref{fig:LAE_complex_z73}), let alone to the LAE itself, suggesting in this case significant ionising photon escape is required from a number of sources that are not included in our (incomplete) spectroscopically confirmed sample, or from fainter sources still. Indeed, an independent photometric selection of galaxy candidates points out a highly overdense region located to the east of JADES-GS-z7-LA \citep{2023arXiv230605295E}.

Additionally, we note this LAE, among the cluster of $z \sim 7.3$ spectroscopically confirmed galaxies, is situated as one of the furthest along the line of sight (from the perspective of an observer at $z = 0$). When considering the next simplest geometry from a single bubble centred on the LAE, two overlapping ionised spheres of equal size coincidentally aligned towards the observer \citep[e.g.][]{2007ApJ...669..663M}, the required volume of the ionised region (and hence the number of ionising photons) may be $\ssim 5\times$ reduced.\footnote{The radii of the two spheres could be three (four) times smaller than in the single-bubble scenario if the LAE is located in the centre (at the far edge) of the sphere furthest along the line of sight, leading to an effective volume decrease by a factor of $9/2$ ($12/2$).} In the specific case of JADES-GS-z7-LA, where there are indications for such chance alignment based on the distribution of nearby galaxies (\cref{fig:LAE_complex_z73}), this could be a viable explanation for the apparent disproportionately large required ionised bubble size $R_\text{ion}^\text{req}$ compared to the estimated ionising photon production rate of neighbouring sources.

In the redshift regime of $6 < z < 7$, we do not find a large number of FRESCO sources surrounding any of the LAEs in our sample. This is perhaps not that surprising: galaxies at $z \sim 6.6$ fall in the `redshift desert' of FRESCO, where \Halpha\ is redshifted out of the F444W grism wavelength coverage (approximately ranging from $3 \, \mathrm{\upmu m}$ to $5 \, \mathrm{\upmu m}$, depending on the location on the detector), while the strong $\OIII \, \lambda \, 4960, 5008 \, \mathrm{\AA}$ and \Hbeta\ lines are not yet contained within it \citep{2023MNRAS.525.2864O}. Additionally, the decreasing sensitivity past a wavelength of $\ssim 4 \, \mathrm{\upmu m}$ and reduced effective survey area of the NIRCam grism imply it is more challenging to find galaxies around a redshift of $z \sim 6.2$-$6.3$. Finally, since FRESCO relies on strong emission-line signatures for spectroscopic confirmation, it is by construction only sensitive to galaxies undergoing bursts of star formation, implying that a significant fraction of sources may be missed if star formation histories are bursty \citep[as suggested by recent findings; e.g.][]{2023arXiv230602470L, 2023arXiv230214155L, 2023arXiv230605295E}. However, if all LAEs are located in extreme overdensities we would still expect to observe a higher number of FRESCO sources.

To more systematically investigate whether LAEs can plausibly create their own bubbles, or whether the contribution of additional (fainter) sources is required, we present a comparison of predicted to minimum required bubble sizes in \cref{fig:Bubble_size_comparison}. Notably, even in the scenario with a smoothly evolving global neutral hydrogren fraction, we infer local ionised bubbles are required to explain the observed \lya\ properties for eight out of the $17$ LAEs considered here. This provides strong evidence for the presence of ionised bubbles at $z > 6$, complementary to the fact that these sources exhibit \lya\ escape fractions comparable to low-redshift LAEs (where IGM absorption does not play a significant role), while having similar rest-frame optical properties \citep[e.g. the \OIII\ to \OII\ line ratio; see][]{2023arXiv230604536S}.

Several of these faint high-redshift LAEs ($z > 6$), with the notable exceptions of ID 4297 and JADES-GS-z7-LA, require relatively small ionised bubbles ($\lesssim 0.3 \, \mathrm{pMpc}$). Particularly considering they represent lower limits, however, the required minimum ionised bubble sizes are still larger than what we estimate the LAEs and their direct neighbours to be able to assemble, assuming the IGM is still fully neutral outside these bubbles ($\bar{x}_\text{\HI} = 1$). The opposite scenario with a smoothly evolving global neutral hydrogen fraction ($\bar{x}_\text{\HI}(z)$) improves the agreement for most of the sources, except for the two highest-redshift LAEs and JADES-GS-z7-LA. We have moreover verified that including photometric galaxy candidates (with limiting magnitude $M_\text{UV} \sim -18 \, \mathrm{mag}$) only marginally increases the number of sources contained within the bubbles, therefore not drastically impacting the predicted bubble size estimates. While it may seem trivial that our incomplete sample of galaxies fails to explain the required sizes of ionised bubbles, this does have important implications: crucially, in either case this implies that the relatively UV-faint LAEs considered here are not solely responsible for carving out their ionised bubbles, unless they are extremely efficient LyC leakers (e.g. if they are able to sustain $f_\text{esc, LyC} \sim 50\%$ for $t_* \sim 50 \, \mathrm{Myr}$, which is at odds with expectations; see \citealt{2023arXiv230604536S}). We note there are indications the same holds true for UV-bright LAEs \citep[e.g.][]{2023arXiv230405385J}. Instead, our results therefore suggest there is still a significant population of undetected, and thus likely ultra-faint ($M_\text{UV} \gtrsim -18 \, \mathrm{mag}$), sources contributing to or even dominating the reionisation of the environments of LAEs in our sample.

\section{Summary and conclusions}
\label{sec:Summary_and_conclusions}

We have investigated the environments of $17$ reionisation-era LAEs identified by \textit{JWST}/NIRSpec as part of JADES observations over GOODS-S \citep{2023arXiv230602471J}. We conservatively estimate sizes of ionised bubbles required to reconcile the measured \lya\ velocity offset with the \lya\ escape fraction \citep{2023arXiv230604536S}. We summarise our main findings as follows:
\begin{itemize}
    \item The relatively low \lya\ velocity offsets ($\Delta v_\text{\lya} \lesssim 300 \, \mathrm{km \, s^{-1}}$) combined with moderately high \lya\ escape fractions ($f_\text{esc, \lya} > 5\%$) observed in our sample of LAEs suggest the presence of ionised bubbles of the order of $R_\text{ion}^\text{req} \sim 0.1$-$1 \, \mathrm{pMpc}$ in a patchy reionisation scenario where the bubbles are still surrounded by a fully neutral IGM. At the highest-redshift regime ($z \gtrsim 6.5$), we find such bubbles are necessitated even if the rest of the IGM is homogeneously (but moderately) reionised.
    \item Around half of the LAEs in our sample are found to coincide with large-scale galaxy overdensities at $z \sim 5.8$-$5.9$ and $z \sim 7.3$ \citep{2023arXiv230605295E, 2023arXiv231104270H}, suggesting \lya\ transmission is strongly enhanced in such regions, and underlining the importance of LAEs as tracers of the first large-scales ionised regions.
    \item Considering only spectroscopically confirmed galaxies, we find our sample of $z > 7$ LAEs and their direct neighbours are generally not able to produce the required ionised bubbles based on the \lya\ transmission properties (assuming ionising radiation escapes from these sources at $f_\text{esc, LyC} = 5\%$ for $t_* = 50 \, \mathrm{Myr}$), suggesting fainter sources ($M_\text{UV} \gtrsim -18 \, \mathrm{mag}$) likely play an important role in carving out these bubbles.
\end{itemize}

We conclude that our findings support the case for faint, numerous star-forming galaxies as the main drivers of cosmic reionization. Harnessing the combined power of NIRSpec multi-slit and NIRCam slitless spectroscopy in acquiring a unique view of the early Universe during cosmic reionisation, these results demonstrate the potential of the most distant LAEs as probes of the reionization. Future observational campaigns with \textit{JWST}, pushing to fainter magnitudes and larger galaxy samples, will therefore undoubtedly help us converge on a more detailed understanding of reionization.

\begin{acknowledgements}
    We thank the anonymous referee for their constructive feedback that helped improve this work. We further thank Callum Witten for helpful suggestions. This work is based on observations made with the NASA/ESA/CSA \textit{James Webb Space Telescope} (\textit{JWST}). The data were obtained from the Mikulski Archive for Space Telescopes at the Space Telescope Science Institute, which is operated by the Association of Universities for Research in Astronomy, Inc., under NASA contract NAS 5-03127 for \textit{JWST}. These observations are associated with programmes 1180, 1210, 1895, and 1963. The authors acknowledge the FRESCO team led by PI Pascal Oesch for developing their observing program with a zero-exclusive-access period. JW, RM, WMB, MC, TJL, LS, and JS acknowledge support by the Science and Technology Facilities Council (STFC), by the European Research Council (ERC) through Advanced Grant 695671, `QUENCH', and by the UK Research and Innovation (UKRI) Frontier Research grant RISEandFALL. JW further gratefully acknowledges support from the Fondation MERAC. RS acknowledges support from an STFC Ernest Rutherford Fellowship (ST/S004831/1). AS, GCJ, AJB, AJC, and JC acknowledge funding from the `FirstGalaxies' Advanced Grant from the ERC under the European Union's Horizon 2020 research and innovation programme (Grant agreement No. 789056). JMH, DJE, BDJ, BER, and CNAW acknowledge a \textit{JWST}/NIRCam contract to the University of Arizona (NAS5-02015). RM also acknowledges support by the STFC and funding from a research professorship from the Royal Society. SA and MP acknowledge support from Grant PID2021-127718NB-I00 funded by the Spanish Ministry of Science and Innovation/State Agency of Research (MICIN/AEI/10.13039/501100011033). This research is supported in part by the Australian Research Council Centre of Excellence for All Sky Astrophysics in 3 Dimensions (ASTRO 3D), through project number CE170100013. SC acknowledges support by European Union's HE ERC Starting Grant 101040227, `WINGS'. ECL acknowledges support of an STFC Webb Fellowship (ST/W001438/1). DJE is supported as a Simons Investigator. MP also acknowledges support from the Programa Atracci\'on de Talento de la Comunidad de Madrid via grant 2018-T2/TIC-11715. H\"U gratefully acknowledges support by the Isaac Newton Trust and by the Kavli Foundation through a Newton-Kavli Junior Fellowship. The research of CCW is supported by NOIRLab, which is managed by the Association of Universities for Research in Astronomy (AURA) under a cooperative agreement with the National Science Foundation. This work has also used the following packages in \program{python}: the \program{SciPy} library \citep{Jones2001}, its packages \program{NumPy} \citep{2011CSE....13b..22V} and \program{Matplotlib} \citep{Hunter2007}, the \program{Astropy} package \citep{2013A&A...558A..33A, 2018AJ....156..123A}.
\end{acknowledgements}


\bibliographystyle{aa} 
\bibliography{Ionised_bubbles}



\appendix

\section{Detailed \texorpdfstring{\lya}{\lyatext} line profile radiative transfer}
\label{app:Line_profile}

In this appendix, we consider the effects of broadened line profiles on the net absorption of \lya\ photons by the IGM. This requires us to assume the full \lya\ spectral profile as it emerges from a galaxy\footnote{In this context, we use intrinsic to refer to the line properties of \lya\ after radiative-transfer processing by the ISM and CGM yet before IGM attenuation.}, which we are prevented from directly observing in the reionisation era. While observations at lower redshift can provide some clues to the expected intrinsic line profiles, the evolving physical properties of galaxies (e.g. specific star formation rate and dust content) may cause them to change as a function of cosmic time \citep{2021ApJ...908...36H}.

\begin{figure}
	\centering
	\includegraphics[width=\linewidth]{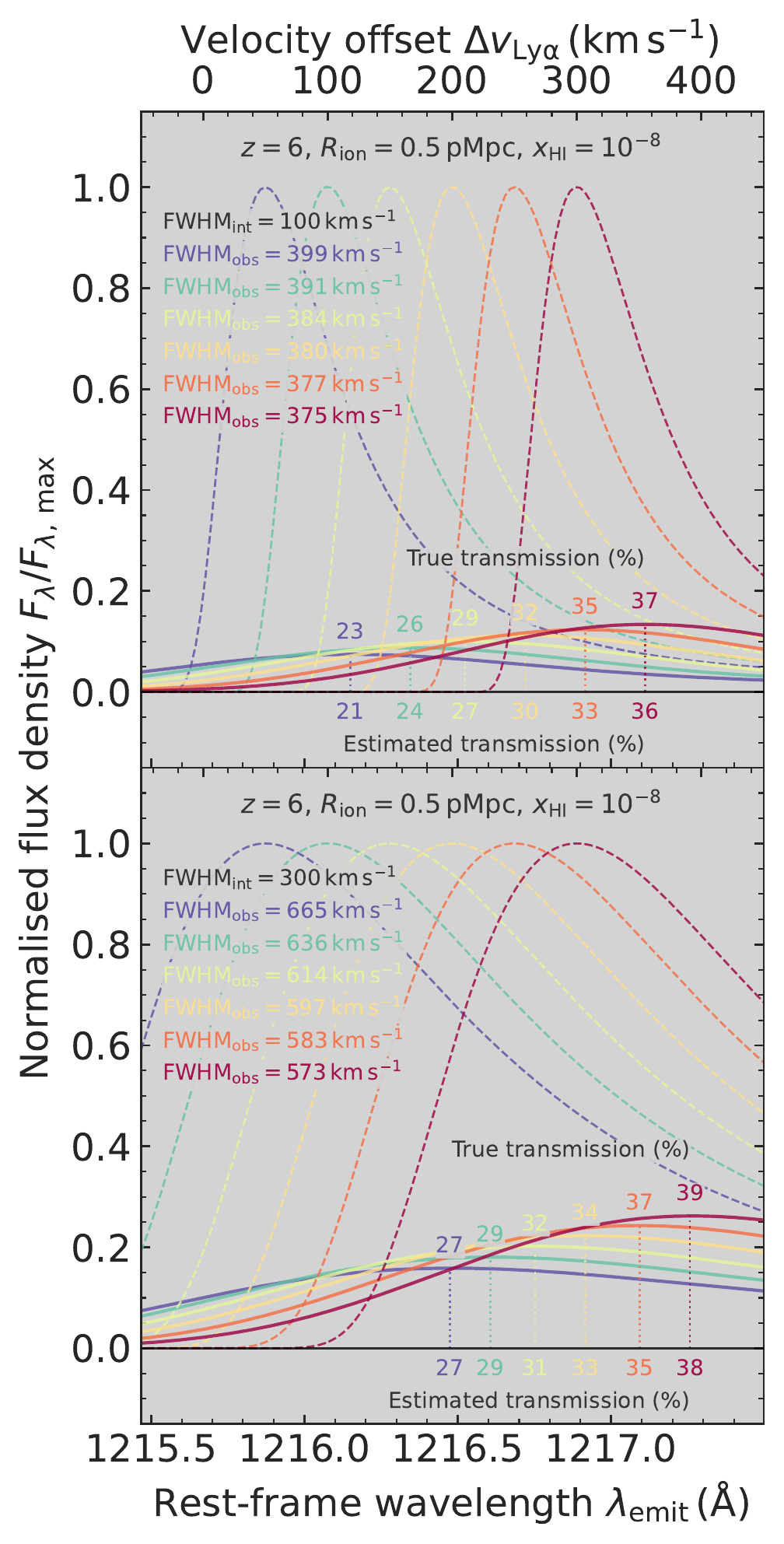}
	\caption{Examples of the wavelength-dependent IGM processing for a source at $z = 6$ situated in an ionised bubble with $R_\text{ion} = 0.5 \, \mathrm{pMpc}$ and $x_\text{\HI} = 10^{-8}$. Intrinsic \lya\ line profiles (dashed coloured lines) are asymmetric Gaussian profiles with varying velocity offsets $\Delta v_\text{\lya}$ and an intrinsic FWHM of $100 \, \mathrm{km \, s^{-1}}$ (top panel) and $300 \, \mathrm{km \, s^{-1}}$ (bottom panel). Observed line profiles (solid coloured lines) are convolved to a fiducial spectral resolution of $R = 1000$. Above the peak of each observed profile (indicated by vertical dotted lines), the true transmission fractions (total flux of the observed line profile divided by the total intrinsic flux) are annotated, while those estimated based on the transmission curve at this wavelength are shown below.}
	\label{fig:Transmission_asym_Gaussian_radiative_transfer}
\end{figure}
\begin{figure*}
	\centering
	\includegraphics[width=\linewidth]{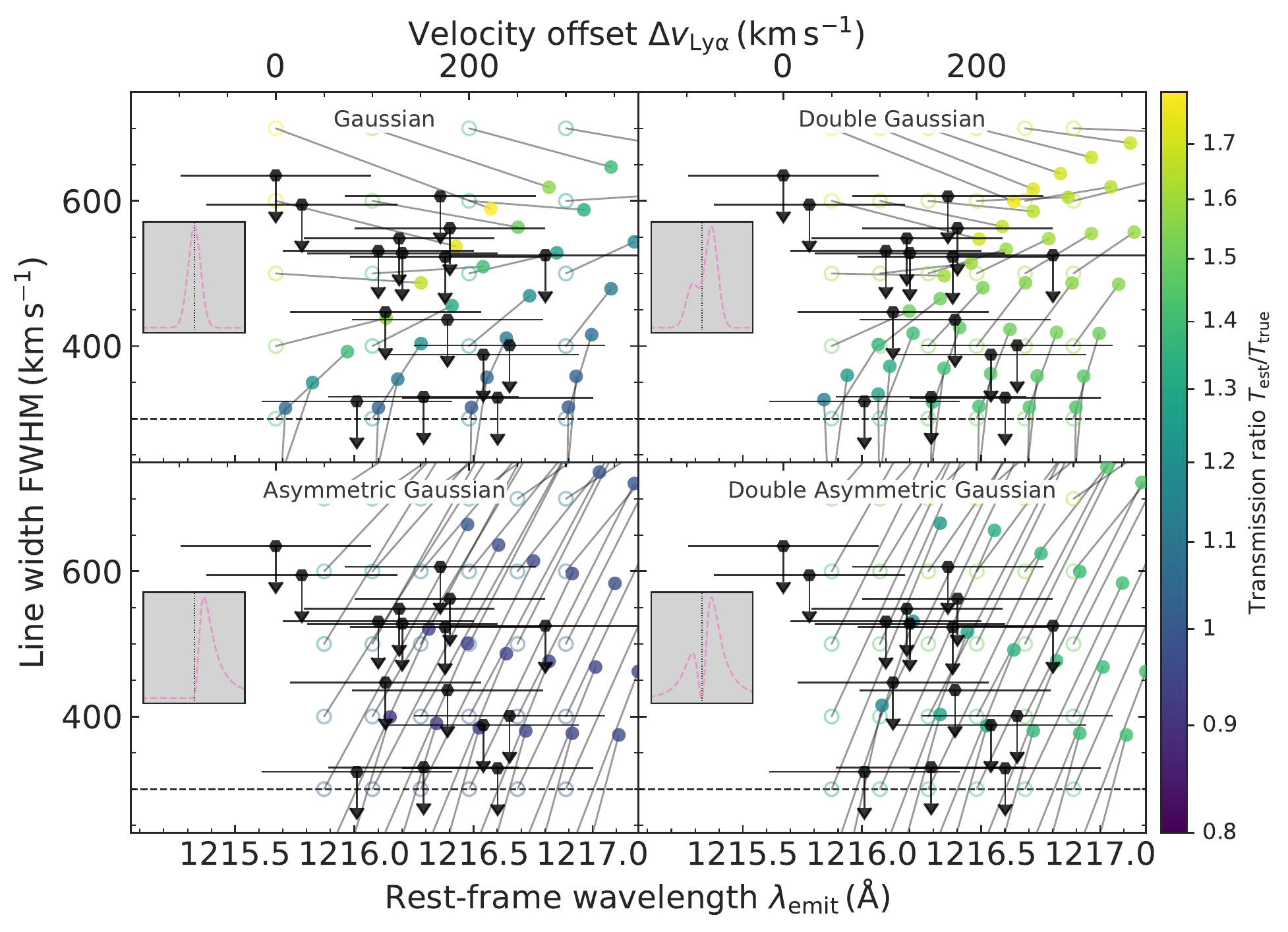}
	\caption{Systematic effects involved in estimating the IGM transmission for different \lya\ line profiles. The intrinsic profiles considered (shown in the grey inset panels) are composed of a single (left columns) or double (with a fraction of $30\%$ and $70\%$ of the total flux contained by the blue and red peaks respectively, which have an equal but opposite velocity offset; right columns) Gaussian profile, either regular (top row) or asymmetric \citep[; bottom row]{2014ApJ...788...74S}. For each intrinsic line profile with a given FWHM and velocity offset $\Delta v_\text{\lya}$ (indicated by open circles), the corresponding observed velocity offset and convolved line width (filled circles) are connected by grey lines. The colour of each point shows the ratio between the estimated and true transmission, $T_\text{est}/T_\text{true}$. A horizontal dashed black line indicates the fiducial spectral resolution of $R = 1000$. Black hexagons show the JADES LAEs (upper limits indicating the line widths have not been deconvolved) considered in this work.}
	\label{fig:Transmission_systematics}
\end{figure*}

Here, we considered several cases to gain insight into the impact of varying the intrinsic \lya\ spectral profile. As a first category, we examined regular Gaussian profiles. Secondly, we considered an empirical asymmetric Gaussian profile \citep{2014ApJ...788...74S}. Motivated by the observed line profile of a galaxy with similar properties as those studied in this work ($M_\text{UV} = -19.6 \, \mathrm{mag}$; $\text{EW}_\text{\lya} \approx 100 \, \mathrm{\AA}$) at $z \simeq 4.88$ \citep{2021A&A...650A..98W}, we choose a fiducial asymmetry parameter of $a_\text{asym} = 0.3$ \citep[see also e.g.][]{2020A&A...635A..82L}. For both categories we considered lines with a given velocity offset consisting either of a single component, or of two components each containing a set fraction of the total flux but at opposite velocity offset). For double-peaked lines, we choose the blue peak to contain a fiducial $30\%$ of the total flux, a conservatively high value based on studies of low-redshift LAEs \citep[e.g.][]{2021ApJ...908...36H}.

We obtained mock observations by attenuating each \lya\ spectral profile as it is assumed to emerge from a galaxy -- a single or double (a)symmetric Gaussian -- by a representative IGM absorption curve. This curve corresponds to the galaxy being centred in an ionised bubble at $z = 6$ with radius $R_\text{ion} = 0.5 \, \mathrm{pMpc}$ and residual neutral fraction of $x_\text{\HI} = 10^{-8}$. Finally, we convolved the spectrum to a fiducial spectral resolution of $R = 1000$. In \cref{fig:Transmission_asym_Gaussian_radiative_transfer}, the case of single asymmetric line profiles is shown as an example, including a comparison between the ratio of observed flux to the intrinsic flux (the `true' transmission) and the transmission at the peak of the observed line profile (which provides our estimate of IGM transmission; \cref{ssec:Ionised_bubble_sizes}). This shows that for the assumed intrinsic line profile, the estimates are consistent with the true transmission, if mildly underestimated. In this case, the size of the ionised bubble would therefore be underestimated by our method.

A more complete assessment of the systematic effects for each assumed intrinsic profile, with varying velocity offset and line width, is shown in \cref{fig:Transmission_systematics}. Given the uncertainty in the precise spectral resolution of our NIRSpec measurements (\cref{ssec:Ionised_bubble_sizes}), we show the convolved line widths of JADES LAEs as upper limits to compare with the mock observations described above. While each simulated line profile is systematically shifted to redder wavelengths due to IGM absorption, the transmission characteristics are entirely dependent on the intrinsic line profile. Since the blue peak experiences near-complete absorption, transmission estimates for the double-peaked profiles are up to a factor $\ssim 2$ higher than the true values, although this only occurs at large velocity offsets ($\Delta v_\text{\lya} \gtrsim 200 \, \mathrm{km \, s^{-1}}$). On the other hand, a single asymmetric red peak consistently has a higher transmission fraction than estimated.

We conclude that the true transmission is highly dependent on the intrinsic line profile. However, it is well approximated by the transmission at the observed peak wavelength for an asymmetric, red-dominated line profile, as commonly seen in low-redshift observations where IGM absorption does not play a major role \citep[e.g.][]{2018ApJ...862L..10E, 2021ApJ...908...36H, 2023ApJ...954L..14H}. In this case, the transmission may even be systematically underestimated by up to $\ssim 20\%$, which again results in effective lower limits on the estimated sizes of ionised bubbles.

%
%

\end{document}